\documentclass[ journal = jctcce, manuscript = article, layout = traditional ]{achemso}

\usepackage[dvipsnames]{xcolor}
\usepackage{chemformula} 
\usepackage{mhchem}      
\usepackage{braket}
\usepackage[T1]{fontenc} 
\usepackage{tikz}
\usepackage{pgfplots}
\usepackage{tikzscale}
\usepackage{lipsum}
\usepackage{pdfpages}

\usetikzlibrary{arrows,chains,matrix,positioning,scopes, fit, calc, decorations, shapes, decorations.pathreplacing, shadows,shadows.blur}
\usetikzlibrary{backgrounds}
\usetikzlibrary{pgfplots.groupplots}
\usetikzlibrary{external}
\tikzsetexternalprefix{plots_compiled/}
\usepgfplotslibrary{colormaps}

\pgfplotsset{
/pgfplots/colormap={jet}{rgb255(0cm)=(0,0,128) rgb255(1cm)=(0,0,255)
rgb255(3cm)=(0,255,255) rgb255(5cm)=(255,255,0) rgb255(7cm)=(255,0,0)
rgb255(8cm)=(128,0,0)}
}

\pgfplotsset{
/pgfplots/colormap={temp}{rgb255=(36,0,217) rgb255=(25,29,247) rgb255=(41,87,255)
rgb255=(61,135,255) rgb255=(87,176,255) rgb255=(117,211,255) rgb255=(153,235,255)
rgb255=(189,249,255) rgb255=(255,255,255) rgb255=(255,255,255) rgb255=(255,242,189)
rgb255=(255,214,153) rgb255=(255,172,117) rgb255=(255,120,87) rgb255=(255,61,61)
rgb255=(247,40,54) rgb255=(217,22,48) rgb255=(166,0,33)}
}

\usepackage{soul}
\usepackage{xcolor}

\usepackage{tikz}
\usetikzlibrary{positioning}
\usetikzlibrary{arrows,shapes}
\usepackage{calc}
\usepackage{forloop}
\usetikzlibrary{calc}
\usepackage{csquotes}

\author{Lars-Hendrik Frahm}
\email{lfrahm@physnet.uni-hamburg.de}
\author{Daniela Pfannkuche}
\affiliation[University of Hamburg]{I. Institut f{\"u}r Theoretische Physik, Universit\"at Hamburg, Jungiusstra{\ss}e 9, 20355 Hamburg, Germany}

\title[Ultrafast ab-initio Quantum Chemistry Using Matrix Product States]{Ultrafast ab-initio Quantum Chemistry Using Matrix Product States}

\begin{document}
   
\begin{abstract}
   Ultrafast dynamics in chemical systems provide a unique access to fundamental processes at the molecular scale. A proper description of such systems is often very challenging because of the quantum nature of the problem. The concept of matrix product states (MPS), however, has proven its performance in describing such correlated quantum system in recent years for a wide range of applications. In this work, we continue the development of the MPS approach to study ultrafast electron dynamics in quantum chemical systems. The method combines time evolution schemes, such as fourth-order Runge-Kutta and Krylov space time evolution, with MPS, in order to solve the time-dependent Schr\"odinger equation efficiently. This allows for describing electron dynamics in molecules on a full configurational interaction (CI) level for a few femtoseconds after excitation. As a benchmark, we compare MPS based calculations to full CI calculations for a chain of hydrogen atoms and for the water molecule. Krylov space time evolution is in particular suited for the MPS approach, as it provides a wide range of opportunities to be adjusted to the reduced MPS dimension case. Finally, we apply the MPS approach to describe charge migration effects in iodoacetylene and find direct agreement between our results and experimental observations.
\end{abstract}

   \section{Introduction}
Ultrafast dynamics lay the foundation in understanding essential processes in chemical and biological systems. Processes such as the formation of chemical bonds, transfer of charges in photovoltaic systems, and photochromism take place on the time scale of femtoseconds\cite{doi:10.1351/pac200072122219,Falke1001,doi:10.1021/cr9800816}. Recent advances in ultrafast science allow to control and observe these processes with a resolution of tens of attoseconds\cite{doi:10.1021/acs.chemrev.6b00453} and thereby to obtain unique insights into the mechanisms at the fast end of the atomic time scale. Due to their light weight, the electrons respond to excitations first, in a mechanism called \textit{charge migration}, which then triggers dynamics involving the nuclei as well, referred to as \textit{charge transfer}. The electron and the nuclei motion are closely coupled to each other. For example, the displacement of the electrons may have a significant impact on the reaction trajectory of the nuclei\cite{jortner:ratner:molecular_electronics}. Having a profound understanding of the electronic dynamics is therefore crucial when investigating how bonds form or when designing artificial reactions\cite{C3CP50591J}. Close collaboration between experiment and theory will be very important to gain the necessary understanding of electron dynamics on the attoseconds time scale. However, the large number of degrees of freedom, in connection with the emerging electronic correlations, makes a quantum theoretical description extremely difficult. 

Various first principle methods have been developed to describe ground states and time-dependent phenomena of molecular systems. The \textit{post-Hartree-Fock methods}\cite{doi:10.1002/9780470125823.ch2,DAVIDSHERRILL1999143,ROOS1980157,doi:10.1063/1.443164,doi:10.1063/1.3193710} start from an uncorrelated (Hartree-Fock) description of the electrons, resulting in molecular orbitals that serve as single electron basis for the many-electron state. One of those methods that approximate the many-electron state and that is particularly successful in describing correlated systems is the \textit{density matrix renormalization group}\cite{PhysRevLett.69.2863,PhysRevB.48.10345} (DMRG). DMRG has mathematically proven advantages when studying one-dimensional systems with short-range interaction and a gapped ground state\cite{1742-5468-2007-08-P08024}. Therefore DMRG quickly became a very popular approach to study phenomena in one dimension, ranging from various ground state phase diagrams \cite{PhysRevB.52.6581,PhysRevLett.89.236401,PhysRevLett.99.216403,PhysRevB.58.6414,PhysRevLett.113.027201,PhysRevLett.80.845,PhysRevB.58.2635} to finite temperature and/or periodic boundary situations\cite{PhysRevLett.93.207204,PhysRevLett.98.070201,PhysRevB.72.220401,PhysRevLett.75.3537,PhysRevB.56.5061,PhysRevB.59.12184}. Even though the mathematically proven advantages are limited to one-dimensional systems, DMRG has been adapted to two \cite{PhysRevB.61.6320,PhysRevLett.109.067201} and higher dimensions \cite{Verstraete:2004cf}. The scaling of entanglement is the main limiting factor for the DMRG method in two and higher dimensions, however, in special situations, DMRG performs well even in multidimensional situations{\cite{PhysRevB.92.235116}}. With its generalization to tensor networks\cite{ORUS2014117}, a variety of other algorithms evolved, ranging from procedures to determine dynamical quantities\cite{PhysRevB.60.335,PhysRevB.66.045114} such as Green's functions or real-time evolution\cite{PhysRevLett.93.076401,1742-5468-2004-04-P04005} up to quantum information theory interpretations\cite{PhysRevLett.93.227205,PhysRevB.70.205118} and applications in machine learning\cite{NIPS2016_6211,2016arXiv160503795N}. 


DMRG has also been applied successfully to quantum chemical problems{\cite{doi:10.1002/qua.24898}} allowing to calculate benchmark level ground states of many molecules\cite{doi:10.1063/1.478295,doi:10.1080/0026897031000155625,doi:10.1146/annurev-physchem-032210-103338,Wouters2014}. By exploiting abelian and non-abelian\cite{0295-5075-57-6-852,PhysRevB.78.245109} symmetries, the efficiency of DMRG has been further improved. DMRG is a very popular tool in many quantum chemistry codes {\cite{MOLPRO-WIREs,doi:10.1021/acs.jctc.7b00174,doi:10.1002/jcc.24221}} today, although, in contrast to one-dimensional and short-ranged problems, there is no guarantee it is able find the numerically exact ground state in polynomial time. Still, some chemical problems may require non-polynomial time to be solved exactly on a quantum mechanical level.\cite{C2CP42695A} Inserting the DMRG solver into other methods of quantum chemistry such as complete active space{\cite{doi:10.1021/ct400707k}} and density matrix embedding theory{\cite{doi:10.1021/acs.jctc.6b00316}}, led to the family of \textit{post-DMRG} methods. Further has DMRG been successfully combined with perturbation theory{\cite{doi:10.1021/ct400707k}} and the coupled cluster approach{\cite{legezaccdmrg}}. In these extensions, DMRG is used as a solver if the resulting active space is too large to perform the necessary operations exactly. Most of these studies focus on ground states, however, also low-lying excited states\cite{doi:10.1080/0026897031000155625} and Green's functions\cite{doi:10.1021/acs.jctc.7b00682} have been investigated using DMRG.

This work serves to continue the development of DMRG based algorithms in the field of quantum chemistry. We want to utilize matrix product states (MPS), the mathematical heart of DMRG, as an efficient representation of the many-electron state and benchmark them as a possible approach to study ultrafast electron dynamics in molecules (charge migration). The dynamic representation of the effective Hilbert space makes MPS in particular suited for the short time response of a molecule to excitation, for example to photoionization. We want to see how MPS compare to calculations including the complete many-electron state and for what period of time we can track the quasi exact full configurational interaction (CI) dynamics. This work is organized as follows: First, we introduce the concept of MPS as efficient representation of the many-electron state of a molecule. We discuss the connection to other post-Hartree-Fock methods and outline the advantages of MPS. We briefly explain how to operate on MPS efficiently, using a variational approach similar to ground state DMRG. We further continue by presenting the time evolution methods used in this work, namely the fourth order Runge-Kutta method and Krylov space time evolution. Second, we benchmark MPS based time evolution with full CI calculations for two distinct molecules. To adapt to existing literature, we study a chain of hydrogen atoms, where we compare the single particle Green's function in both, the real-time and frequency domain. We continue the benchmark of the MPS approach using different flavors of the Krylov space method at the example of the singly ionized water molecule. We discuss resulting features and interpret the results in the context of MPS. Third, we apply the MPS approach to study charge migration in iodoacetylene (\ce{C2HI}). We investigate the ultrafast electron dynamics as a response to an ionization of the iodine atom. The resulting partial charges show the large capability of MPS in quantum chemistry, as we find good agreement between our results and existing experimental and theoretical work. In the last chapter, we summarize and propose further potentials of MPS based time evolution approaches in quantum chemistry.

   \section{Methods}
\subsection{Matrix Product States}
The time-dependent many-electron state of a molecule can be expanded in the occupation number representation of a basis of $L$ orthonormal orbitals
\begin{equation}
\ket{\psi(t)} = \sum_{n_{1\uparrow} n_{1\downarrow } \cdots n_{L\uparrow } n_{L\downarrow }} c_{n_{1\uparrow} n_{1\downarrow } \cdots n_{L\uparrow } n_{L\downarrow }}(t) \ket{n_{1\uparrow} n_{1\downarrow } \cdots n_{L\uparrow } n_{L\downarrow }}, \label{eq:fci}
\end{equation}
where $n_{i\uparrow} \in \{0, 1\}$ ($n_{i\downarrow} \in \{0, 1\}$) is the number of up (down) electrons in the orbital $i$, the coefficient tensor $c_{n_{1\uparrow} n_{1\downarrow } \cdots n_{L\uparrow } n_{L\downarrow }}(t)$ holds the time-dependent expansion coefficients, and $\ket{n_{1\uparrow} n_{1\downarrow } \cdots n_{L\uparrow } n_{L\downarrow }}$ are the occupation number basis states we call configurations in the following. The factorially growing number of possible configurations ${2L\choose {N}}$ makes working with $\ket{\psi(t)}$ so challenging, as it requires to store the same number of coefficients in the expansion (the curse of dimensionality). With latest developments in parallel computing, this many-electron state approach has been handled for up to $22$ orbitals holding $22$ electrons in ground state calculations\cite{doi:10.1063/1.4989858}. However, to study real-time evolution in a reasonable amount of time, the orbital sets need to be much smaller.


For larger sets of orbitals, approximations allow to reduce the number of configurations to a manageable amount. A popular approach is to start from an uncorrelated state (usually the Hartree-Fock ground state $\ket{HF}$) and expand the time-dependent state in terms of electronic excitation classes
\begin{equation}
   \ket{\psi(t)} \approx c_0(t) \ket{HF} + \underbrace{\sum_{ij\sigma} c_{ij}(t) \hat c^{\dagger}_{j\sigma} \hat c_{i\sigma} \ket{HF}}_{\text{1h\ 1p}} + \cdots.
\end{equation}
The $n$-hole $n$-particle class consists of configurations where $n$ electrons have been promoted from a core or valence orbital into one of the virtual orbitals that were unoccupied in the Hartree-Fock ground state. Using a limited number of excitation classes reduces the number of configurations considered and therefore allows to extend the calculations to larger sets of orbitals. These calculations are then referred to as configurational interaction\cite{doi:10.1063/1.1999636} (CI) with, for example, including single excitation only\cite{article} (CIS) or including single and double excitations\cite{PhysRevA.95.053411} (CISD). If all possible classes are included, the resulting many-electron state equals the expansion in Eq. \ref{eq:fci}, which is then referred to as full CI.

Alternatively, in the multi-configurational Hartree-Fock approach\cite{MEYER199073,BECK20001,PhysRevA.83.063416}, one fixes the number of configurations considered in the approximated representation of the exact many-electron state in Eq. \ref{eq:fci}, but optimizes the orbitals represented by those configurations dynamically. The configurations remain fixed throughout the time evolution, however the shape of the orbitals they represent adapts to the time evolution.

These two approaches may have a problem when investigating time-dependent phenomena. In both examples, the number of configurations is restricted artificially, which neglects a big number of configurations that may become important during the time evolution. It is often difficult to tell if the used approach is able to represent the many-electron state appropriately, especially when the underlying mechanisms are unknown and the shape of the necessary configurations changes within the time evolution.

An approach that avoids to limit the configurations in the many-electron state is a wave function decomposition based on matrix product states\cite{SCHOLLWOCK201196,doi:10.1080/14789940801912366} (MPS). MPS have successfully proven to represent ground states in quantum chemistry\cite{Wouters2014,doi:10.1146/annurev-physchem-032210-103338,doi:10.1063/1.478295} and are now on the rise to tackle dynamical quantities such as Green's functions\cite{doi:10.1021/acs.jctc.7b00682}. Here, the coefficient tensor in Eq. \ref{eq:fci} is decomposed into a sequence of matrix products
\begin{equation}
\ket{\psi(t)}_{MPS} = \sum_{n_{1\uparrow} n_{1\downarrow } \cdots n_{L\uparrow } n_{L\downarrow }} A^{n_{1\uparrow}n_{1\downarrow}}(t) A^{n_{2\uparrow}n_{2\downarrow}}(t) \cdots A^{n_{L\uparrow}n_{L\downarrow}}(t) \ket{n_{1\uparrow} n_{1\downarrow } \cdots n_{L\uparrow } n_{L\downarrow }}, \label{eq:mps}
\end{equation}
where the $A^{n_{i\uparrow}n_{i\downarrow}}(t)$ are the time-dependent decomposition matrices and their product represents the coefficient tensor $A^{n_{1\uparrow}n_{1\downarrow}}(t) A^{n_{2\uparrow}n_{2\downarrow}}(t) \cdots A^{n_{L\uparrow}n_{L\downarrow}}(t) = c_{n_{1\uparrow} n_{1\downarrow } \cdots n_{L\uparrow } n_{L\downarrow }}(t)$. There must exist such a decomposition, since we can reshape the coefficient tensor into a matrix and sequentially decompose this matrix using singular value decomposition. Schollw\"ock\cite{SCHOLLWOCK201196} has layed down this concept in his seminal work very detailed.

The factorially growing number of coefficients in Eq. \ref{eq:fci} is now encoded in the rank $D_{FCI}$ of the decomposition matrices $A^{n_{i\uparrow}n_{i\downarrow}}(t)$. There is no computational advantage compared to handling the coefficient tensor in Eq. \ref{eq:fci} at this point. However, the idea of MPS is now to find smaller matrices $\tilde A^{n_{i\uparrow}n_{i\downarrow}}(t)$ that represent the decomposition matrices with little error. The smaller matrices can be constructed from the $D$ largest singular values of $A^{n_{i\uparrow}n_{i\downarrow}}(t)$, reducing the exponential scaling matrix dimension to some fixed value $D$. The dimension $D$ of the reduced matrices will be called the MPS \textit{bond dimension} in the following. This procedure allows to reduce the dimension of the decomposition matrices, while the sequence of reduced matrix products $\tilde A^{n_{1\uparrow}n_{1\downarrow}}(t) \tilde A^{n_{2\uparrow}n_{2\downarrow}}(t) \cdots \tilde A^{n_{L\uparrow}n_{L\downarrow}}(t) \approx c_{n_{1\uparrow} n_{1\downarrow } \cdots n_{L\uparrow } n_{L\downarrow }}(t)$ still gives a quasi optimal\cite{doi:10.1137/090752286} representation of the coefficient tensor. The singular value spectrum decays quickly in most physical situations, especially when the entanglement in $\ket{\psi(t)}$ is limited or short-ranged\cite{SCHOLLWOCK201196}. The actual matrix products are never performed in efficient MPS implementations, but operations are exclusively done on individual matrices\cite{PhysRevB.91.155115} or pairs of matrices when using two-site algorithms\cite{PhysRevLett.69.2863}.


\subsection{Time Evolution Methods}
Our goal is now to find the dynamics of the many-electron state using the more efficient MPS representation. The time evolution is still described by the Schr\"odinger equation for $\ket{\psi(t)} \rightarrow \ket{\psi(t)}_{MPS}$
\begin{equation}
i\hbar \frac{\partial}{\partial t} \ket{\psi(t)}_{MPS} = \hat H \ket{\psi(t)}_{MPS}, \label{eq:sg}
\end{equation}
with the formal solution for time-independent Hamiltonians
\begin{align}
   \ket{\psi(t-t_0)}_{MPS} = e^{-\frac{i}{\hbar}\hat H (t-t_0)} \ket{\psi(t_0)}_{MPS}, \label{eq:time_evo_opera}
\end{align}
and the quantum chemistry Hamiltonian in the Born-Oppenheimer approximation
\begin{align}
   \hat H = \sum_{ij\sigma} t_{ij} \hat c^{\dagger}_{i\sigma}\hat c_{j\sigma} + \frac{1}{2}\sum_{ijkl\sigma\tau} V_{ijkl} \hat c^{\dagger}_{i\sigma}\hat c^{\dagger}_{j\tau}\hat c_{l\tau}\hat c_{k\sigma} + E_0. \label{eq:hamil}
\end{align}
The coefficients $t_{ij}$ represent the kinetic energy of the electrons, as well as, the electron-nuclei interaction in the orbital basis, while the coefficients $V_{ijkl}$ account for the electron-electron interaction. These coefficients can be obtained from any quantum chemistry software\cite{MOLPRO-WIREs, doi:10.1002/jcc.540141112, doi:10.1002/wcms.1340} for a chosen orbital set\cite{doi:10.1021/ci600510j}. Nuclei remain fixed in our study, therefore their kinetic energy vanishes and the nucleus-nucleus interaction appears as the constant energy $E_0$.

Various methods have been developed to solve the time-dependent Schr\"odinger equation in Eq. \ref{eq:sg} using the MPS approach\cite{PhysRevB.72.020404,PhysRevLett.93.040502,PhysRevB.94.165116,1742-5468-2004-04-P04005,1367-2630-8-12-305,PhysRevLett.93.207204,PhysRevB.70.121302,doi:10.1063/1.2080353,PhysRevB.91.165112}. On the one hand, there are methods developed specifically for MPS, which operate on the individual matrices and use properties special to MPS\cite{PhysRevB.91.165112,PhysRevB.94.165116,PhysRevLett.98.070201}. Some of these methods work best or are restricted to problems where the interaction is short-ranged. However, the scattering integrals $t_{ij}$ and $V_{ijkl}$ in the Hamiltonian Eq. \ref{eq:hamil} couple orbitals that sit apart in the one-dimensional list of orbitals, making the quantum chemistry Hamiltonian effectively long-ranged. This forbids the use of time evolving block decimation\cite{PhysRevLett.98.070201}, one of the most popular time evolution methods used for MPS. On the other hand, we can understand the MPS approach as a technique that changes the form we store many-electron states in our computer and how we operate using them. Having efficient procedures to operate on MPS, such as adding MPS or applying the Hamiltonian, we can use any time evolution method that has been developed to solve differential equations like the time-dependent Schr\"odinger equation in Eq. \ref{eq:hamil}. In this work, we want to apply two of those general methods, that we modify to fit calculations using MPS, namely the fourth-order Runge-Kutta method\cite{doi:10.1137/S0036142994260872} as well as Krylov space time evolution\cite{doi:10.1063/1.451548}. But first we explain how we perform the necessary operations on MPS efficiently.

\subsubsection{Operations on MPS} \label{sec:operations}
We need to perform three basic operations when utilizing the Runge-Kutta method and the Krylov space time evolution in order to solve the time-dependent Schr\"odinger equation: scaling MPS by a factor $\alpha\ket{A}_{MPS}$, adding two or more MPS $\ket{A}_{MPS} + \ket{B}_{MPS}$, and applying the Hamiltonian to the MPS, $\hat H\ket{\psi}_{MPS}$.

Scaling an MPS is straight-forward by multiplying all elements of one of the decomposition matrices in Eq. \ref{eq:mps}. When dealing with MPS of large norm we have observed better numerical stability if multiplying all matrices in the MPS by $\sqrt[L]{\alpha}$.

The addition of MPS and the application of an operator on an MPS are more challenging operations, which is due to the higher bond dimension of the decomposition matrices in the resulting MPS. For example, when adding two MPS
\begin{align}
   \ket{C}_{MPS} = \ket{A}_{MPS} + \ket{B}_{MPS}, \label{eq:sum_exact}
\end{align}
where $\ket{A}_{MPS}$ is constructed from matrices with maximum bond dimension $D_A$, $\ket{B}_{MPS}$ is constructed from matrices with maximum bond dimension $D_B$, the resulting MPS $\ket{C}_{MPS}$ is constructed from matrices of bond dimension up to $D_A + D_B$. The matrices in the resulting MPS need to be truncated in a following step to assure the MPS bond dimension does not increase with every following addition. This behavior becomes even more critical when adding many MPS or when applying the Hamiltonian (especially in quantum chemistry with the effective long-range interaction). Intermediate MPS of large bond dimension arise and additional truncation steps increase computational complexity.

To circumvent this problem, we use a variational approach\cite{Verstraete:2004cf,1367-2630-12-5-055026}, which is similar to the ground state procedure in DMRG. Instead of minimizing the energy functional, we minimize the residual norm of an initially random MPS and the sum of MPS $\ket{A}_{MPS} + \ket{B}_{MPS}$
\begin{eqnarray}
   \mathcal{L}[\ket{C}_{MPS}] &=& \left | \left |\ket{C}_{MPS} - ( \ket{A}_{MPS} + \ket{B}_{MPS} ) \right |\right |^2, \label{eq:add_min}
\end{eqnarray}
where $\mathcal{L}[\ket{C}_{MPS}]$ is the Lagrangian whose global minimum is the solution of the MPS sum. We find it by differentiating Eq. \ref{eq:add_min} with respect to the decomposition matrices $C^{n_{i\uparrow}n_{i\downarrow}}$ in $\ket{C}_{MPS}$ and solve the resulting linear system of equations. Then we differentiate with respect to the next decomposition matrix $C^{n_{i+1\uparrow}n_{i+1\downarrow}}$ and continue optimizing $\ket{C}_{MPS}$ to fit Eq. \ref{eq:sum_exact}. This is analogous to the sweeps in regular ground state calculation of DMRG, however instead of solving an eigenvalue equation, we obtain a linear equation, which can be solved easily. After $\approx 10$ sweeps, the matrices in $\ket{C}_{MPS}$ have been optimized to represent $\ket{A}_{MPS} + \ket{B}_{MPS}$, however, under the condition that the decomposition matrices of $\ket{C}_{MPS}$ do not exceed a previously fixed maximum bond dimension $D$. Therefore, we are adding and truncating in one step, which reduces the computational time. This approach also allows to add more than two MPS at once and to apply operators efficiently. Especially when adding more than two MPS or when adding MPS and at the same time applying an operator ($\hat H \ket{A}_{MPS} + \ket{B}_{MPS}$), the resulting MPS is optimized to represent the complete result in the given MPS space. Thus, multiple subsequent truncation steps are avoided, which otherwise could spoil the result of the total operation.

The computational cost of such a sweep is small compared to a DMRG ground state calculation sweep. In ground state DMRG, the algorithm spends most of the time in the repeated operation of the effective Hamiltonian to solve the eigenvalue equation, whereas we just have to solve a linear equation. The computational complexity per sweep reduces to $\mathcal{O}(L^4D^2 + L^3D^3)$ for applying an operator\cite{Wouters2014} and to $\mathcal{O}(LD^3)$ for adding MPS. When working with MPS of limited bond dimensions, adding MPS is simple compared to application of the Hamiltonian. We will keep the number of operator applications as low as possible in the following, whereas the total calculation time is barely affected by performing MPS additions. We have now the necessary tools to handle MPS efficiently and continue to the time evolution methods we will adapt to the MPS approach. 

\subsubsection{Fourth-Order Runge-Kutta}
A popular method to solve a differential equation like Eq. \ref{eq:sg} is to use the fourth-order Runge-Kutta (RK4) method. We expand the time evolved state in terms of the initial state and four Runge-Kutta vectors:
\begin{eqnarray}
\ket{k_1} &:=& \hbar \Delta t \hat H \ket{\psi(t)}\\
\ket{k_2} &:=& \hbar \Delta t \hat H \left (\ket{\psi(t)} + \frac{1}{2}\ket{k_1} \right )\\
\ket{k_3} &:=& \hbar \Delta t \hat H \left (\ket{\psi(t)} + \frac{1}{2}\ket{k_2} \right )\\
\ket{k_4} &:=& \hbar \Delta t \hat H \left (\ket{\psi(t)} + \ket{k_3} \right ).
\end{eqnarray}
The state evolved after a time step $\Delta t$ is then constructed as a superposition of the initial state and the four Runge-Kutta vectors
\begin{equation}
\ket{\psi(t+\Delta t)} = \ket{\psi(t)} + \frac{1}{6}\left( \ket{k_1} + 2 \left ( \ket{k_2} + \ket{k_3} \right) + \ket{k_4} \right ) + \mathcal{O}(\Delta t^5). \label{eq:rk_add}
\end{equation}
In total it takes four Hamiltonian applications and four MPS additions (counting the sum in Eq. \ref{eq:rk_add} as single summation) to perform a time step by $\Delta t$, where all operations can be done efficiently using the operations outlined in Section \ref{sec:operations}. The RK4 method equals the fourth-order Taylor expansion of the time evolution operator in Eq. \ref{eq:time_evo_opera}. Higher order Runge-Kutta methods exist, however, the RK4 methods is widely distributed and has a very small time step error of $\mathcal{O}(\Delta t^5)$. The RK4 method is neither energy conserving nor norm conserving, although, if $\Delta t$ is chosen small, the norm and the energy remain close to constant on the discussed time scales.

\subsubsection{Krylov Space Time Evolution} \label{sec:krylov}
The other method we want to utilize for the MPS approach is a time evolution based on the Krylov space\cite{doi:10.1063/1.451548}. The idea of Krylov space methods is to find a smaller vector space (the Krylov space) that holds the starting vector as well as low orders of $\hat H^n \ket{\psi(t)}$, namely in our MPS case the space 
\begin{align}
\mathcal{K} = span(\{\ket{\psi(t)}_{MPS},\hat H\ket{\psi(t)}_{MPS}, \hat H^2\ket{\psi(t)}_{MPS}, \hat H^3\ket{\psi(t)}_{MPS}\cdots\}). \label{eq:krylov_space}
\end{align} 
The time evolution is then performed in this smaller vector space and the time evolved state is constructed from its basis states.

We will use two different schemes to construct the basis states of the Krylov space: First option is to construct the Krylov space directly as given in Eq. \ref{eq:krylov_space} by sequentially applying the Hamiltonian and normalizing accordingly
\begin{equation}
\ket{\phi^{k+1}}_{MPS} = \frac{\hat H\ket{\phi^{k}}_{MPS}}{|_{MPS}\bra{\phi^k}\hat H \hat H\ket{\phi^{k}}_{MPS}|^2}, \label{eq:kry_non}
\end{equation}
where the first Krylov vector is the state at time $t$, $\ket{\phi^0}_{MPS} = \ket{\psi(t)}_{MPS}$. This basis is non-orthogonal, which has advantages and disadvantages we will be discussing in Section \ref{sec:h2o}.

Second option is to span the Krylov space from a basis that is orthogonalized in a Gram-Schmidt fashion within the construction
\begin{align}
\ket{\phi^{k+1}}_{MPS} = \hat H \ket{\phi^{k}}_{MPS} - \sum_{j \leq k} \frac{_{MPS}\bra{\phi^j} \hat H\ket{\phi^k}_{MPS}}{_{MPS}\braket{\phi^j | \phi^{j}}_{MPS}} \ket{\phi^j}_{MPS}. \label{eq:kry_orth}
\end{align}
This creates an orthogonal basis for the Krylov space in Eq. \ref{eq:krylov_space}, however, this basis has a couple of disadvantages when working with truncated MPS, which we will also discuss in Section \ref{sec:h2o}. The procedure to generate the Krylov basis initially proposed by Lanczos requires the orthogonalization only between the new, $\phi_{k+1}$, and the two previous Krylov states $\phi_{k-1}$, $\phi_{k}${\cite{Lanczos:1950zz}. The orthogonality to all remaining Krylov states is automatically given. In our case, with regard to the truncated MPS representation of the Krylov states, we perform a full orthogonalization with respect to all Krylov vectors to counteract truncation and numerical errors.}

The state evolved after a time step $\Delta t$ is constructed from the Krylov basis states by building the superposition
\begin{equation}
\ket{\psi(t + \Delta t)}_{MPS} = \sum_{k}\left[ e^{-i\frac{\Delta t}{\hbar} N^{-1} H} \right]_{k0} \ket{\phi^k}_{MPS} \label{eq:krylov_evo}
\end{equation}
where the matrix $N_{ij} = _{MPS}\braket{\phi_i | \phi_j}_{MPS}$ represents the overlaps of the Krylov basis states, $H_{ij} = _{MPS}\braket{\phi_i | \hat H | \phi_j}_{MPS}$ represents the Hamiltonian in the Krylov space and $[ \  \cdot \  ]_{k0}$ denotes the element at the $k$-th row and the $0$-th column of the exponential matrix. The extra $N^{-1}$ in the exponential takes the possible non-orthogonality of the Krylov basis vectors into account and corrects the matrix exponential accordingly.{\cite{doi:10.1063/1.451548, 1367-2630-8-12-305}} The time evolution operator in Eq. {\ref{eq:krylov_evo}} follows from the Schr\"odinger equation for state expansions in non-orthogonal bases $i \hbar \dot c_k = \sum_{k'} (N^{-1} H)_{kk'} c_{k'}$, for which the condition of unitarity $\frac{\partial}{\partial t} \braket{\psi|\psi} = 0$ can be easily proven. Therefore, if the time evolution is performed in the Krylov space, this method is intrinsically norm and energy conserving.

In our time-dependent study, we noticed convergence of the dynamics on the full CI level using Krylov spaces of four to six basis states and time step sizes around $\Delta t \sim 1as$.  When using a Krylov space dimension of $N_{K}$ basis vectors, we need $N_{K} - 1$ Hamiltonian applications and one summation of MPS to build the superposition in Eq. \ref{eq:krylov_evo}. If using the orthonormal construction, there are $N_{K} - 1$ additional summations. A Krylov space dimension of $N_{K} = 5$ is similar to the RK4 method in computational effort.

In principle, both schemes to generate the Krylov space (the vector space generated from vectors in Eq. {\ref{eq:kry_non}} and the vector space generated from vectors Eq. {\ref{eq:kry_orth}}) should match and exactly span the Krylov space in Eq. {\ref{eq:krylov_space}}. However, in case of truncated MPS, there might be situations where one approach is superior to the other. If the first Krylov vector (the state to evolve $\ket{\psi(t)}_{MPS}$) is close to an eigenvector of the Hamiltonian, using the orthogonalized Krylov vectors from Eq. {\ref{eq:kry_orth}} is more stable. In this case, the non-orthogonalized Krylov vectors tend to be linearly dependent, causing the overlap matrix $N$ to approaches singularity, and the inverse in the exponential of Eq. {\ref{eq:krylov_evo}} to diverge. This spoils the results of the time evolution calculation. Otherwise, if the initial state is far from an eigenstate of the Hamiltonian, the vector space generated from non-orthogonalized vectors span the Krylov space more correctly, since they use the limited degrees of freedom of the truncated MPS basis more efficiently. This effect will be observed and discussed in detail in Section {\ref{sec:h2o}}.

Krylov time evolution has a couple of advantages compared to other methods for time-dependent MPS. First, it is norm and energy conserving outrunning the RK4 method outlined above. Minor changes to the norm and the energy will occur due to the truncation of the MPS when performing the superposition in Eq. \ref{eq:krylov_evo}, however, that is not due to the time evolution method, but is related to the finite bond dimension of the matrices in the MPS. For comparison, in the RK4 method state norm is lost due to both, the non-unitary time evolution as well as the truncation of the MPS when constructing the state in Eq. \ref{eq:rk_add}. Second, the Krylov space method allows for better adjustment to MPS with limited bond dimension. We will see in Section \ref{sec:h2o} that the performance of the MPS approach depends on the form of the Krylov basis states and the size of the Krylov space dimension. The sole systematic error of our Krylov+MPS approach is caused by the small number of Krylov vectors, the time step size, and the truncation of the MPS. The first two sources will be easily controlled with respect to the MPS size, however, even with MPS of small bond dimension we can approach full CI like results.

   \section{Results}
Now, we compare the MPS representation of the many-electron state from Eq. \ref{eq:mps} to the full CI state representation in Eq. \ref{eq:fci} using both, the RK4 and the Krylov time evolution method. We start with a chain of ten hydrogen atoms \ce{H10} in a STO-6G basis and the water molecule \ce{H2O} in a 6-31G basis. These orbital sets are small enough to calculate full CI results in a reasonable amount of time. 

Later, we will apply the MPS approach to study charge migration effects in iodoacetylene using a number of molecular orbitals that is too large to be treated on a full CI level. Here, we will see how MPS find the necessary Hilbert space dynamically and allow for proper prediction of the electron dynamics following an ionization.

\subsection{Chain of Hydrogen Atoms} \label{sec:h10}
We start the benchmark by comparing MPS and full CI results for a system of ten hydrogen atoms arranged on a chain. Atomic orbitals are described on the STO-6G basis set level, resulting in a single $1s$ orbital per atom. This gives us a model of $10$ electrons distributed over $10$ orbitals; a system size easily treatable applying the full CI approach. Its simple one-dimensional structure and the single orbital per atom basis set, makes this model very similar to the one-dimensional Hubbard model, although it is slightly more complex due to the inter-orbital long-ranged Coulomb interaction and electron hopping between atoms that are apart on the chain. Thanks to the one-dimensionality of the problem, the hydrogen chain\cite{doi:10.1063/1.2345196,doi:10.1063/1.1449459,doi:10.1002/qua.23173,doi:10.1021/acs.jctc.7b00682}, as well as the Hubbard model\cite{PhysRevB.52.6581,PhysRevLett.89.236401,PhysRevLett.99.216403,PhysRevB.58.6414,PhysRevLett.113.027201,PhysRevLett.80.845,PhysRevB.58.2635}, already gained much attention in the literature, exploring the MPS approach in strongly correlated situations. These studies also have been extended to multi-orbital sites{\cite{PhysRevB.90.245129}}.

We adapt a situation motivated by Ronca et al.\cite{doi:10.1021/acs.jctc.7b00682}, who are to our knowledge the first that applied the MPS approach to study dynamical quantities in ab-initio quantum chemistry. The hydrogen atoms have the equilibrium bond distance of $0.95 \text{\normalfont\AA} $ and we use the basis of orthonormalized atomic orbitals throughout the calculation. The L\"owdin-orthonormalized atomic orbitals and the one- and two-particle matrix elements in Eq. \ref{eq:hamil} were obtained from the open-source quantum chemistry package PySCF\cite{doi:10.1002/wcms.1340}.

First quantity of comparison is the spin summed and normalized one-particle Green's function in the time domain
\begin{align}
G_{ij} (t - t') = - i \Theta(t-t') \frac{\sum_{\sigma}\phantom{1}_{MPS}\braket{ \psi_0 | c^{\dagger}_{i\sigma} e^{\frac{i}{\hbar}\hat H (t -t')} c_{j\sigma} | \psi_0}_{MPS}}{\sum_{\sigma} \phantom{1}_{MPS} \braket{ \psi_0 | c^{\dagger}_{i\sigma} c_{j\sigma} | \psi_0}_{MPS} }, \label{eq:green_time}
\end{align}
with the system at time $(t-t') = 0$ being in the MPS representation of the ground state $\ket{\psi_0}_{MPS}$. We can understand the Green's function as the response of the system to a sudden annihilation of an electron in orbital $j$ and the probability to find an electron move to orbital $i$ after the time $(t - t')$ as a result of the annihilation process. Applying the time propagation methods explained in the previous section for backward propagation, we have the theoretical framework to approach this quantity efficiently.

\begin{figure}
   \centering
   \begin{tikzpicture}
   \begin{groupplot}[
      group style={
          group name=my plots,
          group size=1 by 2,
          ylabels at=edge left,
          vertical sep=10pt
      },
      footnotesize,
      width=229pt,
      height=100pt,
      tickpos=left,
      ytick align=outside,
      xtick align=outside,
      enlarge x limits=false 
  ]
  
  \nextgroupplot[
         xlabel={},
         xlabel near ticks,
         xtick={0.0,1.0,2.0,3.0,4.0},
         xticklabels={},
         xmin=-0.1, 
         xmax= 4.1,
         ylabel={$G^{\text{Im}}_{55}(t-t')$} ,
         xmajorgrids, 
         ymajorgrids, 
         ytick={-1.00,0.0, 1.00},
         yticklabels={$-1.0$,$0.0$,$1.0$},
         ymin=-1.1,
         ymax=1.1,
         after end axis/.code={     \draw[line width = 1.5pt, color = gray!15] 
                                 (rel axis cs:0,0)rectangle(rel axis cs:1,1);
                                 \draw[line width = 1.5pt,rounded corners=3pt, color = black!80] 
                                 (rel axis cs:0,0)rectangle(rel axis cs:1,1);
                                    },
         legend entries={\scriptsize KRY-FCI,\scriptsize KRY-MPS, \scriptsize RK4-MPS},
         legend columns=3,
         legend style={rounded corners=3pt, line width = 1.5pt,at={(0.5,1.4)},anchor=north,font=\small},                                    
     ]

     \addplot[color=blue, line width = 1pt] table [x expr=\thisrowno{0} * 0.02418884254, y expr=\thisrowno{2} ] {plots/h10_dynamics/dynFCI.out};
     \addplot[color=red, line width = 1pt] table [x expr=\thisrowno{0} * 0.02418884254, y expr=\thisrowno{2} ] {plots/h10_dynamics/dynKRY.out};
     \addplot[color=green, line width = 1pt] table [x expr=\thisrowno{0} * 0.02418884254, y expr=\thisrowno{2} ] {plots/h10_dynamics/dynRK.out};
  \nextgroupplot[
         xlabel=$t-t'$,
         xlabel near ticks,
         xtick={0.0,1.0,2.0,3.0,4.0},
         xticklabels={$0.0fs$,$1.0fs$,$2.0fs$, $3.0fs$, $4.0fs$},
         xmin=-0.1, 
         xmax= 4.1,
         xmajorgrids, 
         ylabel= {abs. error},
         ymajorgrids, 
         ytick={0.0,0.10, 0.20, 0.30},
         yticklabels={$0.0$,$0.1$,$0.2$,$0.3$},
         ymin=-0.03,
         ymax=0.33,
         after end axis/.code={
                                 \draw[line width = 1.5pt, color = gray!15] 
                                 (rel axis cs:0,0)rectangle(rel axis cs:1,1);
                                 \draw[line width = 1.50000pt,rounded corners=3pt, color = black!80] 
                                 (rel axis cs:0,0)rectangle(rel axis cs:1,1);
                                     }
         ]
         \addplot[color=red, line width = 1pt] table [x expr=\thisrowno{0} * 0.02418884254, y expr=\thisrowno{1} ] {plots/h10_dynamics/diffKRY.out};
         \addplot[color=green, line width = 1pt] table [x expr=\thisrowno{0} * 0.02418884254, y expr=\thisrowno{1} ] {plots/h10_dynamics/diffRK.out};
  \end{groupplot}
\end{tikzpicture}
   \caption{(top) The imaginary part of the one-particle Green's function for annihilation and creation of an electron at the fifth hydrogen atom, calculated using the MPS approach, as well as the full CI approach. The maximum MPS bond dimension is $D=30$ and time step size for all calculations is $\Delta t = 1.21 as$. The KRY-MPS result and the KRY-FCI result use a Krylov space dimension of five ($N_{K} = 5$). (bottom) The absolute error of the one-particle Green's function calculated using the MPS approaches compared to the full CI approach.}
   \label{fig:green_time_diff}
\end{figure}
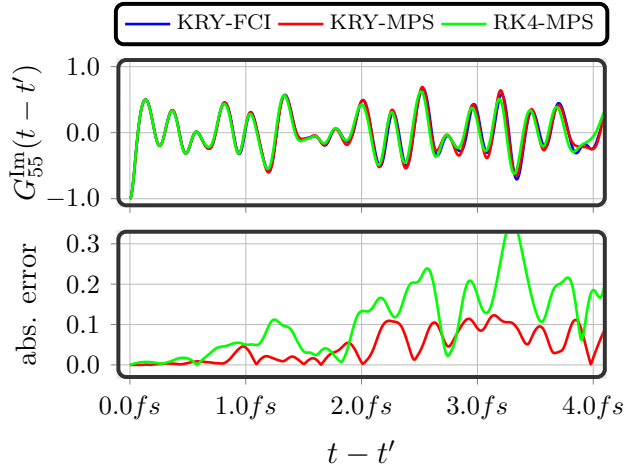
First step in order to calculate the one-particle Green's function in Eq. \ref{eq:green_time} is to find the MPS representation of the ground state $\ket{\psi_0}_{MPS}$. The full CI decomposition of the coefficient tensor in Eq. \ref{eq:fci} results in an MPS with maximum bond dimension $D_{FCI} = 462$ in our spin adapted MPS code, which is based on the open-source code CheMPS2 by Wouters et al.\cite{CheMPS2cite1, CheMPS2cite2, CheMPS2cite3, CheMPS2cite4}. An MPS bond dimension of $D_{FCI} = 462$ is easily handleable in a ground state calculation, therefore we can employ an almost full CI initial state $\ket{\psi_0}_{MPS}$ to avoid effects from truncation of the initial state in our benchmark. To obtain the Green's function, we then annihilate an electron at site $j$ from the initial state ($c_{j\sigma} \ket{ \psi_0}_{MPS}$), evolve the ionized state backwards in time ($e^{\frac{i}{\hbar}\hat H (t -t')}$) and project the time evolved state onto the initial state with an applied excitation at site $i$ $(_{MPS}\bra{ \psi_0} c^{\dagger}_{i\sigma}$).

Fig. \ref{fig:green_time_diff} (top) shows the imaginary part of the one-particle Green's function for an annihilation and creation of an electron at the fifth atom of the hydrogen chain, changing within the first four femtoseconds. The one-particle Green's function was calculated using three different procedures: 
\begin{itemize}
   \item (KRY-FCI) A quasi exact calculation using the full CI setting from Eq. \ref{eq:fci} to represent the many-electron state and the Krylov space method with orthonormal basis states from Eq. \ref{eq:kry_orth} to solve the time-dependent Schr\"odinger equation. The Krylov space dimension is $N_K = 5$ and the time step size is $\Delta t = 1.21 as$, which is well converged (see Supporting Information).
   \item (KRY-MPS) A calculation using the MPS approach from Eq. \ref{eq:mps} to represent the many-electron state and the Krylov space methods with orthonormal basis states from Eq. \ref{eq:kry_orth} to solve the time-dependent Schr\"odinger equation. The maximum bond dimension of the MPS is $D=30$, the Krylov space dimension is $N_K = 5$ and the time step size is $\Delta t = 1.21 as$.
   \item (RK4-MPS) A calculation using the MPS approach from Eq. \ref{eq:mps} to represent the many-electron state and the RK4 method from Eq. \ref{eq:rk_add} to solve the time-dependent Schr\"odinger equation. The maximum bond dimension of the MPS is $D=30$ and the time step size is $\Delta t = 1.21 as$.
\end{itemize}
Both MPS calculations are comparable in calculation time, as both methods need most of time for the four applications of the Hamiltonian in Eq. \ref{eq:kry_orth} and Eq. \ref{eq:rk_add}. We observe, the quasi exact KRY-FCI result is almost completely shadowed by the KRY-MPS result, showing the good correspondence of the KRY-MPS to the quasi exact result. The RK4-MPS result starts to deviate from the KRY-FCI result at about two femtoseconds, however, it still covers the major frequencies in this time frame. In Fig. \ref{fig:green_time_diff} (bottom) we see, how the KRY-MPS result matches the exact result within a deviation of up to $0.1$ from the exact result, whereas the RK4-MPS result acquires an error of up to $0.3$.

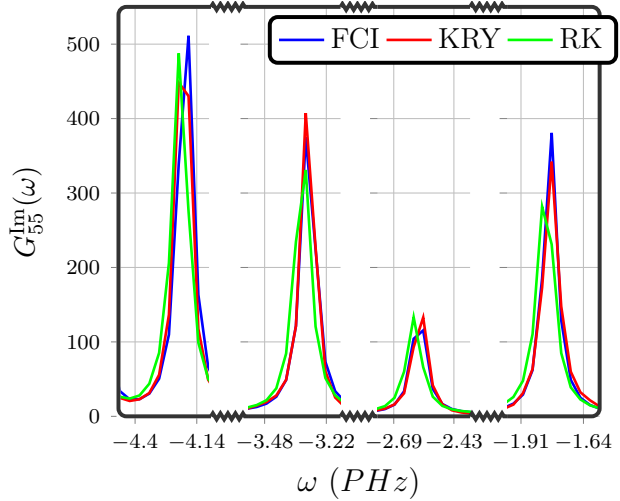
\begin{figure}
   \centering
   \begin{tikzpicture}
   \begin{groupplot}[
      group style={
          group name=my plots,
          group size=4 by 1,
          ylabels at=edge left,
          horizontal sep=14pt
      },
      footnotesize,
      width=80pt,
      height=200pt,
      tickpos=left,
      ytick align=outside,
      xtick align=outside,
      enlarge x limits=false 
  ]
  
  \nextgroupplot[
         xlabel={},
         xlabel near ticks,
         xtick={-4.40,-4.14},
         ticklabel style = {font=\scriptsize} ,
         xmin=-4.47, 
         xmax=-4.08,
         xmajorgrids, 
         ymajorgrids, 
         ytick={0.00, 100, 200, 300, 400, 500},
         ymin=-0.002,
         ymax=550.055,
         ylabel={$G^{\text{Im}}_{55}(\omega)$},
         scaled y ticks = false,
         after end axis/.code={     \draw[line width = 1.5pt, color = white] 
                                 (rel axis cs:0,0)rectangle(rel axis cs:1,1);
                                 \draw[line width = 1.5pt,rounded corners=3pt, color = black!80] 
                                 (rel axis cs:1,0) -- (rel axis cs:0,0) -- (rel axis cs:0,1) -- (rel axis cs:1,1);
                                    },
     ]

     \addplot[color=blue, line width = 1pt] table [x expr=6.57969*\thisrowno{0}, y expr=-\thisrowno{2} ] {plots/h10_green/resultFCI.out};
     \addplot[color=red, line width = 1pt] table [x expr=6.57969*\thisrowno{0}, y expr=-\thisrowno{2} ] {plots/h10_green/resultKRY.out};
     \addplot[color=green, line width = 1pt] table [x expr=6.57969*\thisrowno{0}, y expr=-\thisrowno{2} ] {plots/h10_green/resultRK.out};
   \nextgroupplot[
      xlabel={},
      xlabel near ticks,
      xtick={-3.48,-3.22},
      ticklabel style = {font=\scriptsize},
      xmin=-3.55, 
      xmax=-3.16,
      ylabel={},
      xmajorgrids, 
      ymajorgrids, 
      ytick={0.00, 100, 200, 300, 400, 500},
      yticklabels={},
      ymin=-0.002,
      ymax=550.055,
      axis line style={draw=none},
      scaled y ticks = false,
      after end axis/.code={  
                              \draw[line width = 1.5pt,rounded corners=3pt, color = black!80] 
                              (rel axis cs:0,0) -- (rel axis cs:1,0);
                              \draw[line width = 1.5pt,rounded corners=3pt, color = black!80] 
                              (rel axis cs:0,1) -- (rel axis cs:1,1);                              
                                 },
   ]
   \addplot[color=blue, line width = 1pt] table [x expr=6.57969*\thisrowno{0}, y expr=-\thisrowno{2} ] {plots/h10_green/resultFCI.out};
   \addplot[color=red, line width = 1pt] table [x expr=6.57969*\thisrowno{0}, y expr=-\thisrowno{2} ] {plots/h10_green/resultKRY.out};
   \addplot[color=green, line width = 1pt] table [x expr=6.57969*\thisrowno{0}, y expr=-\thisrowno{2} ] {plots/h10_green/resultRK.out};
   
  \nextgroupplot[
   xlabel={},
   xlabel near ticks,
   xtick={-2.69,-2.43},
   ticklabel style = {font=\scriptsize},
   xmin=-2.76, 
   xmax=-2.36,
   ylabel={},
   xmajorgrids, 
   ymajorgrids, 
   ytick={0.00, 100, 200, 300, 400, 500},
   yticklabels={},
   ymin=-0.002,
   ymax=550.055,
   axis line style={draw=none},
   scaled y ticks = false,
   after end axis/.code={  
                           \draw[line width = 1.5pt,rounded corners=3pt, color = black!80] 
                           (rel axis cs:0,0) -- (rel axis cs:1,0);
                           \draw[line width = 1.5pt,rounded corners=3pt, color = black!80] 
                           (rel axis cs:0,1) -- (rel axis cs:1,1);                              
                              },
]

\addplot[color=blue, line width = 1pt] table [x expr=6.57969*\thisrowno{0}, y expr=-\thisrowno{2} ] {plots/h10_green/resultFCI.out};
\addplot[color=red, line width = 1pt] table [x expr=6.57969*\thisrowno{0}, y expr=-\thisrowno{2} ] {plots/h10_green/resultKRY.out};
\addplot[color=green, line width = 1pt] table [x expr=6.57969*\thisrowno{0}, y expr=-\thisrowno{2} ] {plots/h10_green/resultRK.out};
\nextgroupplot[
   xlabel={},
   xlabel near ticks,
   xtick={-1.91,-1.64},
   ticklabel style = {font=\scriptsize},
   xmin=-1.97, 
   xmax=-1.57,
   ylabel={},
   xmajorgrids, 
   ymajorgrids, 
   ytick={0.00, 100, 200, 300, 400, 500},
   yticklabels={},
   ymin=-0.002,
   ymax=550.055,
   axis line style={draw=none},
   scaled y ticks = false,
   after end axis/.code={  
                           \draw[line width = 1.5pt,rounded corners=3pt, color = black!80] 
                           (rel axis cs:0,1) -- (rel axis cs:1,1) -- (rel axis cs:1,0) -- (rel axis cs:0,0)  ;                              
                              },
   legend entries={\small FCI,\small KRY, \small RK},
   legend columns=3,
   legend style={rounded corners=3pt, line width = 1.5pt,at={(1.2000001,0.98)},anchor=north east,font=\small},                                    
]

\addplot[color=blue, line width = 1pt] table [x expr=6.57969*\thisrowno{0}, y expr=-\thisrowno{2} ] {plots/h10_green/resultFCI.out};
\addplot[color=red, line width = 1pt] table [x expr=6.57969*\thisrowno{0}, y expr=-\thisrowno{2} ] {plots/h10_green/resultKRY.out};
\addplot[color=green, line width = 1pt] table [x expr=6.57969*\thisrowno{0}, y expr=-\thisrowno{2} ] {plots/h10_green/resultRK.out}; 
\end{groupplot}
\draw[line width = 1.5pt, color = black!80, decoration = {zigzag,segment length = 1mm, amplitude = 0.4mm},decorate] (my plots c1r1.south east)--(my plots c2r1.south west);
\draw[line width = 1.5pt, color = black!80, decoration = {zigzag,segment length = 1mm, amplitude = 0.4mm},decorate] (my plots c1r1.north east)--(my plots c2r1.north west);
\draw[line width = 1.5pt, color = black!80, decoration = {zigzag,segment length = 1mm, amplitude = 0.4mm},decorate] (my plots c2r1.south east)--(my plots c3r1.south west);
\draw[line width = 1.5pt, color = black!80, decoration = {zigzag,segment length = 1mm, amplitude = 0.4mm},decorate] (my plots c2r1.north east)--(my plots c3r1.north west);
\draw[line width = 1.5pt, color = black!80, decoration = {zigzag,segment length = 1mm, amplitude = 0.4mm},decorate] (my plots c3r1.south east)--(my plots c4r1.south west);
\draw[line width = 1.5pt, color = black!80, decoration = {zigzag,segment length = 1mm, amplitude = 0.4mm},decorate] (my plots c3r1.north east)--(my plots c4r1.north west);
\node[anchor=north, yshift=-15pt] at ($(my plots c1r1.south west)!0.5!(my plots c4r1.south east)$){$\omega$ ($PHz$)};
\end{tikzpicture}
   \caption{The major peaks of the imaginary part of the frequency dependent one-particle Green's function for a chain of ten hydrogen atoms. All curves are derived from a Fourier transform of time-dependent one-particle Green's function evaluated up to $24.18fs$ using a time step size of $\Delta t = 1.21 as$. The maximum dimension for the MPS approaches is $D=30$ and the FCI and KRY approaches use a Krylov dimension of $N_K=5$. A broadening of $\eta = 0.032 PHz$ has been applied to extract the major peaks. }
   \label{fig:green_frequency}
\end{figure}
When obtaining the one-particle Green's function in the frequencies domain
\begin{align}
   G_{ij} (\omega) = \int^{\infty}_{-\infty} d(t-t')e^{i\omega(t-t')} G_{ij} (t - t'),
\end{align}
from the Fourier transform of Eq. {\ref{eq:green_time}} the MPS results need to be correct over an extended period of time. Therefore, the KRY-MPS approach will perform better here, as it stays close to the full CI one-particle Green's function longer. This can be observed in Fig. \ref{fig:green_frequency}, which shows the major peaks of the one-particle Green's function calculated using the three different approaches introduced before. The total evolution time is $24.18fs$ and a broadening of $\eta = 0.032 PHz$ has been applied to extract the major peaks. We see how the KRY-MPS result peaks almost exactly align with the quasi exact result from KRY-FCI approach. The RK4-MPS result is also close, however, peaks are slightly shifted to smaller frequencies. Neither of our results shows the non-physical behavior of negative one-particle Green's functions that have been reported earlier\cite{doi:10.1021/acs.jctc.7b00682}. We think this is due to our very general approach with every Krylov/Runge-Kutta vector being represented by an MPS in its own optimized virtual basis. This improves the results compared to methods that rely on averaged virtual basis sets\cite{PhysRevB.66.045114,PhysRevB.60.335,doi:10.1021/acs.jctc.7b00682}. Alternatively, the dynamical DMRG method can calculate the one-particle Green's function in the frequency domain directly by minimizing a Hylleraas-like functional\cite{PhysRevB.66.045114}, which may be more efficient if only a specific frequency range is of interest\cite{doi:10.1021/acs.jctc.7b00682}. 

We have seen that our time evolution methods is able to follow the full CI dynamics efficiently using comparably small MPS bond dimension ($D = 30 $ to $D_{FCI} = 462$) and that the Krylov space method allows for longer time evolution. We will continue benchmarking the Krylov method with full CI results, however, we will apply it to situations where the system is higher dimensional and also where choosing the best MPS bond dimension and time evolution parameters is more challenging.

\subsection{\ce{H2O} results } \label{sec:h2o}
In Section \ref{sec:h10}, we have seen how the KRY-MPS approach compares to the RK4-MPS approach when benchmarking with the full CI results. Having almost similar computational demand, the KRY-MPS approach is able to follow the full CI results longer, plus the advantage of being norm and energy conserving. This reduces the risk of non-physical results and makes the approach more stable to MPS bond dimension truncation. Therefore, we focus on the KRY-MPS approach in the following and investigate its stability on the MPS bond dimension $D$, the time step size $\Delta t$, the Krylov space dimension $N_{K}$, as well as the effect of Krylov space basis orthonormalization. As a benchmark molecule, we use the water molecule \ce{H2O} described on a 6-31G basis set level resulting in a basis set of $13$ orbitals holding ten electrons. This is still solvable using the full CI approach, allowing us to perform a profound comparison of our MPS based results and full CI results.

As our ultimate goal is to study charge migration effects in ionized molecules, we proceed with a setting similar to what has been applied in previous work on charge migration effects. Following the idea of correlation driven charge migration we use an uncorrelated state as initial state. Before ionization, the ten electrons of water occupy the five lowest Hartree-Fock molecular orbitals in a completely uncorrelated fashion, leaving eight unoccupied virtual orbitals in our 6-31G basis set. To prepare the initial state, we annihilate one electron in the $1s$ orbital of the oxygen atom. Coherence and correlation will then evolve in the system by the subsequent electron dynamics. This molecular orbital picture has been applied to study correlation driven charge migration before, using static representations of the Hilbert space methods such as CISD\cite{PhysRevA.95.053411}.

The remaining electrons in the molecule start to react to the ionized situation, which results in charge oscillations between the hydrogen atoms and the oxygen atom. We keep the nuclei fixed for simplicity, although they might start moving in realistic situations. These calculations serve to understand the performance of the MPS approach and not to describe the very simple dissociation dynamics in ionized water molecules. We focus on the resulting correlated states and how the MPS approach is able to describe them using small MPS bond dimensions. 

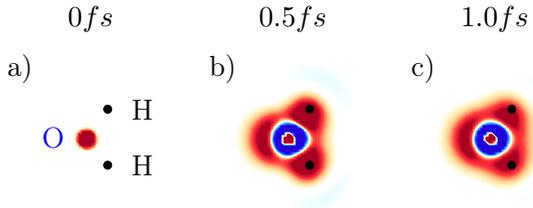
\begin{figure}
   \centering
   \begin{tikzpicture}
   \begin{groupplot}[
      group style={
          group name=my plots,
          group size=3 by 1,
          ylabels at=edge left,
          horizontal sep=5pt,
          vertical sep=30pt
      },
      footnotesize,
      width=135pt,
      tickpos=left,
      ytick align=outside,
      xtick align=outside,
      enlarge x limits=false 
  ]
  \nextgroupplot[
   title = {$0fs$},        
   ymax = 5,
   ymin = -5,
   xmin = -5,
   xmax = 5,
   axis equal image,
   point meta min=-0.05,
   point meta max=0.05,
   hide axis,
   colormap name=temp,
]
   \addplot [matrix plot*,point meta=explicit, shader=interp] file [meta=index 2] {plots/h2o_charge/density_t_000.dat};
   \draw[fill, color=blue] (axis cs:-0.11294891023749,0) circle (0.005pt) node[left, xshift=-5pt] {\small O};
   \draw[fill, color=black] (axis cs:0.89643966569149,1.4837496121996) circle (1.5pt) node[right, xshift=5pt] {\small H};
   \draw[fill, color=black] (axis cs:0.89643966569149,-1.4837496121996) circle (1.5pt) node[right, xshift=5pt] {\small H};
   \node[anchor=north west] at (rel axis cs:0,1) {\small a)};
   \nextgroupplot[
      title = {$0.5fs$},        
      ymax = 5,
      ymin = -5,
      xmin = -5,
      xmax = 5,
      axis equal image,
      point meta min=-0.05,
      point meta max=0.05,
      hide axis,
      colormap name=temp,
   ]
   \addplot [matrix plot*,point meta=explicit, shader=interp] file [meta=index 2] {plots/h2o_charge/density_t_075.dat};
   \draw[fill, color=blue] (axis cs:-0.11294891023749,0) circle (0.005pt);
   \draw[fill, color=black] (axis cs:0.89643966569148,1.4837496121996) circle (1.5pt);
   \draw[fill, color=black] (axis cs:0.89643966569148,-1.4837496121996) circle (1.5pt);
   \node[anchor=north west] at (rel axis cs:0,1) {\small b)};
   
   \nextgroupplot[
      title = {$1.0fs$},        
      ymax = 5,
      ymin = -5,
      xmin = -5,
      xmax = 5,
      axis equal image,
      point meta min=-0.05,
      point meta max=0.05,
      hide axis,
      colormap name=temp,
   ]
   \addplot [matrix plot*,point meta=explicit, shader=interp] file [meta=index 2] {plots/h2o_charge/density_t_178.dat};
   \draw[fill, color=blue] (axis cs:-0.11294891023749,0) circle (0.005pt);
   \draw[fill, color=black] (axis cs:0.89643966569149,1.4837496121996) circle (1.5pt);
   \draw[fill, color=black] (axis cs:0.89643966569148,-1.4837496121996) circle (1.5pt);
   \node[anchor=north west] at (rel axis cs:0,1) {\small c)};
      
   \end{groupplot}
\end{tikzpicture}
   \caption{The time-dependent hole density of a \ce{H2O} molecule with single ionized $1s$ orbital at the oxygen atom. Results derived from a full CI calculation in a 6-31G basis set as an example of the underlying electron dynamics for the following benchmark of the MPS approach.}
   \label{fig:h2o_charge}
\end{figure}
The most important quantity when studying charge migration and investigating the time resolved charge motion is the one-body reduced density matrix (1BRDM)
\begin{align}
   \gamma_{ij} (t) = \sum_{\sigma} \braket{ \psi (t) | c^{\dagger}_{i\sigma} c_{j\sigma} | \psi (t) }.
\end{align}
With the 1BRDM we can calculate the time-dependent charge density 
\begin{align}
   \rho(\mathbf{r}, t) = \sum_{ij} \gamma_{ij}(t) \phi_i(\mathbf{r}) \phi^{*}_j(\mathbf{r}),
\end{align}
where $\phi_i(\mathbf{r})$ are the basis orbitals (here molecular orbitals derived from a Hartree-Fock calculation). We will also use the hole density $\rho^{1h}(\mathbf{r}, t) = \rho^{HF}(\mathbf{r}) - \rho(\mathbf{r}, t)$ that describes the charge difference between the uncorrelated and natural Hartree-Fock ground state $\rho^{HF}(\mathbf{r})$ and the time evolved ionized state. Figure \ref{fig:h2o_charge} shows the charge dynamics with an initially single ionized $1s$ orbital at the oxygen and the following response of the remaining electrons The shown hole densities were calculated using the open source tool ORBKIT\cite{orbkit}.

\begin{figure}
   \centering
   \begin{tikzpicture}
   \begin{groupplot}[
      group style={
          group name=my plots,
          group size=1 by 1,
          ylabels at=edge left,
          vertical sep=10pt
      },
      footnotesize,
      width=229pt,
      height=170pt,
      tickpos=left,
      ytick align=outside,
      xtick align=outside,
      enlarge x limits=false 
  ]
  \nextgroupplot[
         xlabel=$t$,
         xlabel near ticks,
         xtick={0.0,0.25,0.5,0.75,1.0},
         xticklabels={$0.0fs$,$0.25fs$,$0.5fs$, $0.75fs$, $1.0fs$},
         xmin=-0.03, 
         xmax= 1.03,
         xmajorgrids, 
         ylabel= {rel. error $\epsilon(t)$},
         ymajorgrids, 
         ytick={0.0,0.05, 0.10},
         yticklabels={$0\%$,$5\%$,$10\%$},
         ymin=-0.005,
         ymax=0.105,
         after end axis/.code={
                                 \draw[line width = 1.5pt, color = gray!15] 
                                 (rel axis cs:0,0)rectangle(rel axis cs:1,1);
                                 \draw[line width = 1.5pt,rounded corners=3pt, color = black!80] 
                                 (rel axis cs:0,0)rectangle(rel axis cs:1,1);
                                     }
         ]
         \addplot[color=red, line width = 1pt, only marks, mark size=0.4pt] table [x expr=\thisrowno{0} * 0.02418884254, y expr=\thisrowno{1} ] {plots/h2o_dynamics/diffKRY_5_0020_060_non-ortho.out};
         \addplot[color=green, line width = 1pt, only marks, mark size=0.4pt] table [x expr=\thisrowno{0} * 0.02418884254, y expr=\thisrowno{1} ] {plots/h2o_dynamics/diffKRY_5_0020_060_ortho.out};
         \draw[fill, color=black, line width = 1pt, ->] (axis cs:0.55,0.08) -- (axis cs:0.65,0.072) node[line width = 1.0pt,rounded corners=3pt,rectangle,draw,at start, anchor=east, fill=white] {\small orthogonal basis};
         \draw[fill, color=black, line width = 1pt, ->] (axis cs:0.68,0.015) -- (axis cs:0.75,0.035) node[line width = 1.0pt,rounded corners=3pt,rectangle,draw,at start, anchor=north, fill=white] {\small non-orthogonal basis};         
  \end{groupplot}
\end{tikzpicture}
   \caption{The error of the one-body reduced density matrix calculated using the Krylov MPS approach within the first femtosecond. The full CI calculation compared to employs a Krylov space dimension of $N_K = 6$ and a time step size of $\Delta t=0.484 as$. Both MPS calculations use a Krylov space dimension of $N_K = 5$, the same time step size of $\Delta t=0.484 as$ and a maximum MPS bond dimension of $D=60$. }
   \label{fig:h2o_dynamics}
\end{figure}
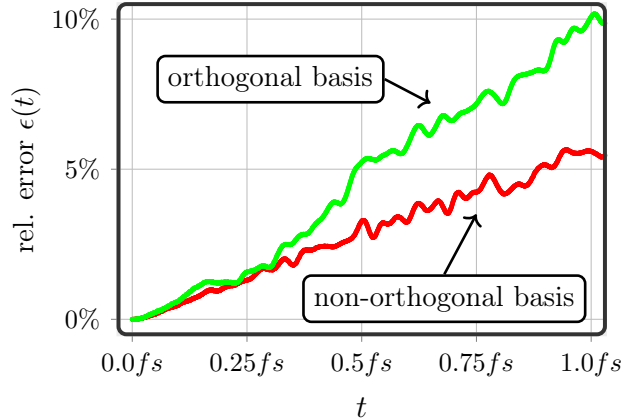
To quantify the precision of our MPS approach, we compare the relative error of the 1BRDM
\begin{align}
   \epsilon(t) = || \gamma^{MPS}(t) - \gamma^{FCI}(t) || / || \gamma^{FCI}(t) ||,
\end{align}
where $\gamma^{MPS}(t)$ is the 1BRDM obtained from the MPS approach and $\gamma^{FCI}(t)$ is the 1BRDM obtained from the quasi exact full CI approach. This error grows at least linearly with time, due to the MPS truncation done, every time we perform a discrete time step. This behavior is known as the run away error of the MPS approach, where the MPS is unable to keep track of the entanglement emerging during the time evolution{\cite{SCHOLLWOCK201196}}. The slope may dependent on the various parameters of the Krylov method and the MPS properties. Figure \ref{fig:h2o_dynamics} demonstrates the different slopes of the error for two MPS calculations, both with Krylov space dimension of $N_K = 5$, similar time step size of $\Delta t=0.484 as$ and a maximum MPS bond dimension of $D=60$, but one using an orthonormalized Krylov basis (green) and the other using non-orthonormalized Krylov space basis vectors (red). The molecular orbitals in the MPS calculation are ordered according to a Fiedler reduction of the band width of the Hamiltonian  matrix\cite{PhysRevA.83.012508}. All full CI calculation to find $\gamma^{FCI}(t)$ in this Section use a conservative Krylov space dimension of $N_K = 6$ and orthonormalized basis states.

The non-orthonormalized Krylov space method has a smaller slope than the approach using orthonormalized Krylov space basis vectors. In Section \ref{sec:krylov}, we explained two options to span the Krylov space: one using orthonormal basis vectors, whereas the other just spans the basis using $H^n \ket{\phi^0}_{MPS}$. Now we see different precision of these options, however, both methods are supposed to span the same Krylov space. The reason for this behavior arises from the limited size of the MPS bond dimension, which effectively restricts the degrees of freedom of the MPS. Whereas the MPS for the non-orthonormal case are fitted to represent $H^n \ket{\phi^0}_{MPS}$ in the optimal way for some given $D$, the orthonormalized states are fitted to represent $H^n \ket{\phi^0}_{MPS}$ and $- \sum_{j \leq k} \frac{_{MPS}\bra{\phi^j}H\ket{\phi^k}_{MPS}}{_{MPS}\braket{\phi^j | \phi^{j}}_{MPS}} \ket{\phi^j}_{MPS}$ on an equal footing (see Eq. \ref{eq:kry_orth}). Therefore, they tend to fulfill the orthonormalization condition, but loose the Krylov space character. The states may be orthogonal to each other, however, they do not span the Krylov space in Eq. {\ref{eq:krylov_space}} correctly. This effect ultimately results in loss of precision when the MPS bond dimension is limited. The Taylor expansion of the time evolution operator in Eq. \ref{eq:time_evo_opera} is closer to the non-orthonormalized Krylov space vectors in Eq. \ref{eq:kry_non} than to the orthonormalized Krylov space vectors calculated in Eq. \ref{eq:kry_orth}. This explains the smaller slope of the error growth, which allows the non-orthonormalized MPS approach to stay around a $5\%$ error after the first femtosecond of time evolution.

\begin{figure}
   \centering
   \begin{tikzpicture}
   \begin{groupplot}[
      group style={
          group name=my plots,
          group size=3 by 2,
          ylabels at=edge left,
          horizontal sep=40pt,
          vertical sep=40pt
      },
      footnotesize,
      width=145pt,
      height=145pt,
      tickpos=left,
      ytick align=outside,
      xtick align=outside,
      enlarge x limits=false 
  ]
  \nextgroupplot[
   title = {$N_{K}=4$ ortho},
   ylabel={$\Delta t$ in $as$},
   xmax = 185,
   xmin = 25,
   ymin = 0.0075,
   ymax = 0.0975,
   point meta min=0.0,
   point meta max=0.3,
   xtick={30, 60, 90, 120, 150, 180},
   xticklabels={$30$,$60$,$90$,$120$,$150$,$180$},
   ytick={0.010335343643938, 0.031006030931813, 0.051676718219689, 0.072347405507564, 0.09301809279544},
   yticklabels={$0.25$, $0.75$, $1.25$, $1.75$, $2.25$},
   scaled y ticks = false,
   colormap name=jet,
 after end axis/.code={
   \draw[line width = 1.25pt, color = white] 
   (rel axis cs:0,0)rectangle(rel axis cs:1,1);
   \draw[line width = 1.25pt,rounded corners=2pt, color = black!80] 
   (rel axis cs:0,0)rectangle(rel axis cs:1,1);
       }
]

\addplot [matrix plot*,point meta=explicit] file [meta=index 2] {plots/h2o_error_kry/err_kry-4-ortho.out};
     \nextgroupplot[
      title = {$N_{K}=5$ ortho},      
      xmax = 185,
      xmin = 25,
      ymin = 0.0075,
      ymax = 0.0975,
      point meta min=0.0,
      point meta max=0.3,
      xtick={30, 60, 90, 120, 150, 180},
      xticklabels={$30$,$60$,$90$,$120$,$150$,$180$},
      ytick={0.010335343643938, 0.031006030931813, 0.051676718219689, 0.072347405507564, 0.09301809279544},
      yticklabels={$0.25$, $0.75$, $1.25$, $1.75$, $2.25$},
      scaled y ticks = false,
      colormap name=jet,
    after end axis/.code={
      \draw[line width = 1.25pt, color = white] 
      (rel axis cs:0,0)rectangle(rel axis cs:1,1);
      \draw[line width = 1.25pt,rounded corners=2pt, color = black!80] 
      (rel axis cs:0,0)rectangle(rel axis cs:1,1);
          },
   ]
   \addplot [matrix plot*,point meta=explicit] file [meta=index 2] {plots/h2o_error_kry/err_kry-5-ortho.out};     
   \nextgroupplot[
      title = {$N_{K}=6$ ortho},    
      xmax = 185,
      xmin = 25,
      ymin = 0.0075,
      ymax = 0.0975,
      point meta min=0.0,
      point meta max=0.3,
      xtick={30, 60, 90, 120, 150, 180},
      xticklabels={$30$,$60$,$90$,$120$,$150$,$180$},
      ytick={0.010335343643938, 0.031006030931813, 0.051676718219689, 0.072347405507564, 0.09301809279544},
      yticklabels={$0.25$, $0.75$, $1.25$, $1.75$, $2.25$},
      scaled y ticks = false,
      colormap name=jet,
    after end axis/.code={
      \draw[line width = 1.25pt, color = white] 
      (rel axis cs:0,0)rectangle(rel axis cs:1,1);
      \draw[line width = 1.25pt,rounded corners=2pt, color = black!80] 
      (rel axis cs:0,0)rectangle(rel axis cs:1,1);
          },
          colorbar right,
          every colorbar/.append style={
             height=2*\pgfkeysvalueof{/pgfplots/parent axis height}+\pgfkeysvalueof{/pgfplots/group/vertical sep},
             ytick={0, 0.1, 0.2, 0.3},
             yticklabels = {$0\%$,$10\%$,$20\%$,$\geq 30\%$}, 
          }
   ]
   \addplot [matrix plot*,point meta=explicit] file [meta=index 2] {plots/h2o_error_kry/err_kry-6-ortho.out}; 
   \nextgroupplot[
      title = {$N_{K}=4$ non-ortho},
      ylabel={$\Delta t$ in $as$},
      xlabel={$D$},
      xmax = 185,
      xmin = 25,
      ymin = 0.0075,
      ymax = 0.0975,
      point meta min=0.0,
      point meta max=0.3,
      xtick={30, 60, 90, 120, 150, 180},
      xticklabels={$30$,$60$,$90$,$120$,$150$,$180$},
      ytick={0.010335343643938, 0.031006030931813, 0.051676718219689, 0.072347405507564, 0.09301809279544},
      yticklabels={$0.25$, $0.75$, $1.25$, $1.75$, $2.25$},
      scaled y ticks = false,
      colormap name=jet,
      after end axis/.code={
          \draw[line width = 1.25pt, color = white] (rel axis cs:0,0)rectangle(rel axis cs:1,1);
          \draw[line width = 1.25pt,rounded corners=2pt, color = black!80] (rel axis cs:0,0)rectangle(rel axis cs:1,1);
       },
       legend to name={CommonLegend}
   ]

   \addplot [matrix plot*,point meta=explicit,] file [y expr=\thisrowno{1}*24.18884254, meta=index 2] {plots/h2o_error_kry/err_kry-4-non-ortho.out};
   \nextgroupplot[
      title = {$N_{K}=5$ non-ortho},
      xlabel={$D$},
      xmax = 185,
      xmin = 25,
      ymin = 0.0075,
      ymax = 0.0975,
      point meta min=0.0,
      point meta max=0.3,
      xtick={30, 60, 90, 120, 150, 180},
      xticklabels={$30$,$60$,$90$,$120$,$150$,$180$},
      ytick={0.010335343643938, 0.031006030931813, 0.051676718219689, 0.072347405507564, 0.09301809279544},
      yticklabels={$0.25$, $0.75$, $1.25$, $1.75$, $2.25$},
      scaled y ticks = false,
      colormap name=jet,
    after end axis/.code={
      \draw[line width = 1.25pt, color = white] 
      (rel axis cs:0,0)rectangle(rel axis cs:1,1);
      \draw[line width = 1.25pt,rounded corners=2pt, color = black!80] 
      (rel axis cs:0,0)rectangle(rel axis cs:1,1);
          },
   ]
   \addplot [matrix plot*,point meta=explicit] file [meta=index 2] {plots/h2o_error_kry/err_kry-5-non-ortho.out};   
 \nextgroupplot[
   title = {$N_{K}=6$ non-ortho},
   xlabel={$D$},
   xmax = 185,
   xmin = 25,
   ymin = 0.0075,
   ymax = 0.0975,
   point meta min=0.0,
   point meta max=0.3,
   xtick={30, 60, 90, 120, 150, 180},
   xticklabels={$30$,$60$,$90$,$120$,$150$,$180$},
   ytick={0.010335343643938, 0.031006030931813, 0.051676718219689, 0.072347405507564, 0.09301809279544},
   yticklabels={$0.25$, $0.75$, $1.25$, $1.75$, $2.25$},
   scaled y ticks = false,
   colormap name=jet,
 after end axis/.code={
   \draw[line width = 1.25pt, color = white] 
   (rel axis cs:0,0)rectangle(rel axis cs:1,1);
   \draw[line width = 1.25pt,rounded corners=2pt, color = black!80] 
   (rel axis cs:0,0)rectangle(rel axis cs:1,1);
       }
]
\addplot [matrix plot*,point meta=explicit] file [meta=index 2] {plots/h2o_error_kry/err_kry-6-non-ortho.out};  
  \end{groupplot}
\end{tikzpicture}
   \caption{The relative error of the 1BRDM for single ionized water molecule $\epsilon(t)$, evaluated one femtosecond after single ionization of the $1s$ orbital. The corresponding full CI calculation employed a Krylov space dimension of $N_K = 6$ and the same time step size $\Delta t$ as given in the figure. }
   \label{fig:oedm_diff}
\end{figure}
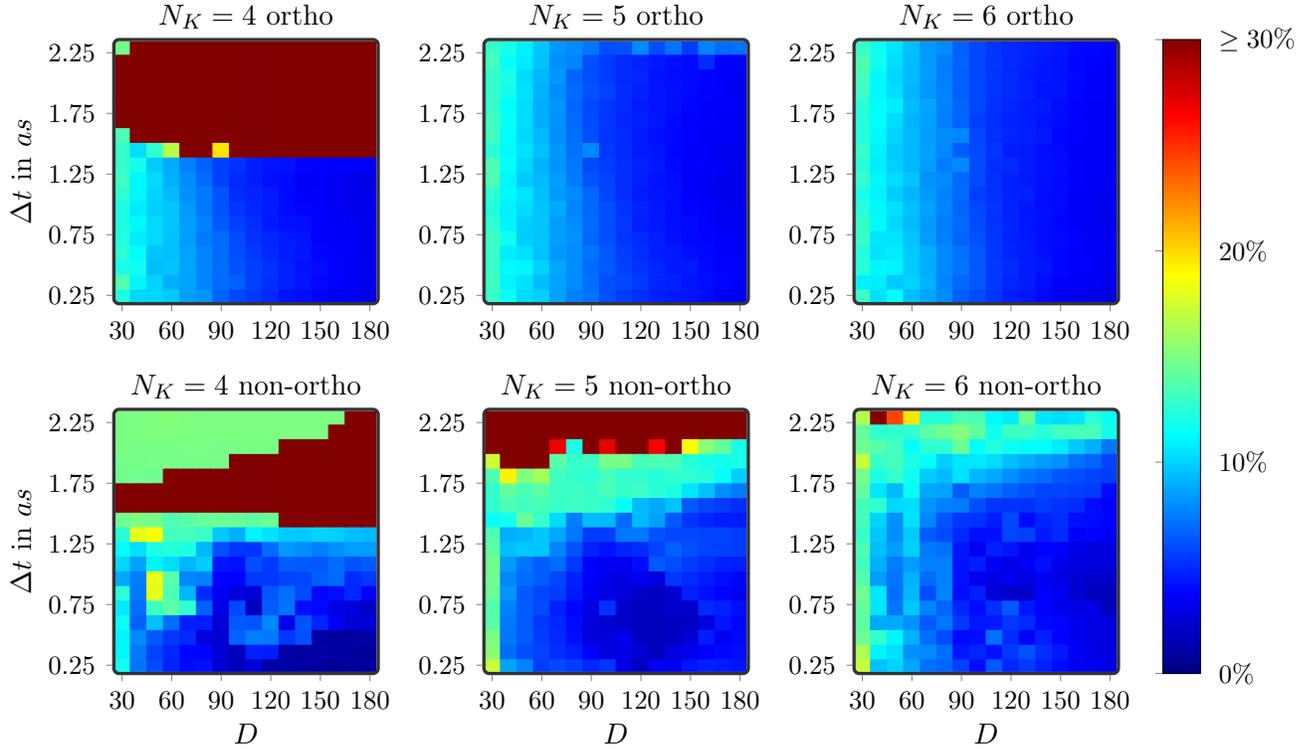
The general dependence of the error of the 1BRDM $\epsilon(t)$ on the MPS bond dimension, on the time step size and on Krylov space orthonormalization can be observed in Figure \ref{fig:oedm_diff}. The error is shown at propagation time $t=1fs$, but averaged over a time frame of $\pm 25as$. This time frame is below the expected Auger decay rate of $\approx 2.1fs$\cite{KESKIRAHKONEN1974139}, so we expect the hole to stay localized in the system. The corresponding full CI results compared to in Figure \ref{fig:oedm_diff} are still obtained from a Krylov space dimension of $N_K = 6$ and the same time step size as the MPS calculation compared to. In the small $\Delta t$ and large bond dimension limit ($\Delta t < 0.045a.u.$ and $D > 100$) all MPS calculations are able to represent the full CI 1BRDM up to an error of $5\%$. Here the results are independent of the Krylov space dimension and the large MPS bond dimension $D$ allows to cover most of the necessary Hilbert space to represent the correlated electron dynamics. This is already a significant reduction in the effective Hilbert space dimension, as the factorization of the coefficient tensor on Eq. \ref{eq:fci} leads to a MPS bond dimension of $D_{FCI} = 622$.

Using non-orthonormalized Krylov basis states allows to further reduce the MPS bond dimension. The bond dimension can be reduced to as little as $D=40$ to achieve an agreement with the full CI result within $5\%$. This requires to reduce the time steps size (here $\Delta t=1.1 as$), due to arising problems with the truncated MPS. When working with non-orthonormal Krylov space basis vectors and small MPS bond dimensions, the Krylov space basis vectors tend to become aligned quickly. The overlap matrix in Eq. \ref{eq:krylov_evo} then approaches singularity and numerical problems arise due to the divergence of the inverse in the matrix exponential in Eq. \ref{eq:krylov_evo}. However this behavior can be easily balanced by using smaller time steps (see Figure \ref{fig:oedm_diff} lower row). Due to the third order scaling of the calculation time with the MPS bond dimension, it is more challenging computationally to handle large MPS bond dimensions than it is to use smaller time steps. We gain a computational advantage, when using the non-orthonormalization Krylov space method and small time steps instead of the orthonormalized method with larger MPS bond dimension and larger time steps. 

The problem with calculating the inverse in Eq. \ref{eq:krylov_evo} also explains why we observe worse convergence when increasing the Krylov space dimension. As we increase the number of Krylov space basis vectors, the closer the overlap matrix in Eq. \ref{eq:krylov_evo} approaches a singular matrix. Then we have the same numerical issues with calculating the inverse as explained above. Therefore, an increase of the Krylov space dimension does not guarantee to improve the results when working with non-orthogonalized basis vectors. Working with smaller Krylov space dimensions and reducing the time step size is the more promising convergence strategy. Summarizing, the non-orthogonalized Krylov space method requires a more careful convergence test, however, the potential reduction in the necessary bond dimension is significant ($33\%$ in this example) and can allow for a description of much more complex situations. If the initial state does not allow for usage of the non-orthogonalized method, we showed orthogonalized Krylov space basis vectors give very robust results and good convergence in terms of time step size and MPS bond dimension.

\subsection{Iodoacetylene}\label{sec:c2hi}
Now we are having a profound understanding of the performance of our MPS approach in realistic conditions and can start to apply the method to situations where full CI calculations are out of modern computational means. We want to use the method to study charge migration in iodoacetylene \ce{C2HI} after photoionization. In a recent experiment by Kraus et al.\cite{Kraus790} they showed very good control over the electron dynamics following ionization by using high harmonic spectroscopy. They are able to track the electronic motion with a time resolution of $100as$. We want to model that situation using our new developed MPS approach and see how the electron dynamics compare to those observed in the experiment.

The electron dynamics observed in the experiment mostly involve valence orbitals and virtual orbitals, which allows us to represent the electrons in the core orbitals in terms of effective core potentials. This reduces the number of electrons and orbitals to a reasonable size that our MPS approach is able to handle efficiently. We employ Stuttgart effective core potentials, removing $46$ electrons from the iodine atom\cite{doi:10.1002/jcc.10037} and two electrons from each carbon atom \cite{doi:10.1080/00268979300103121}. The hydrogen is included on the 6-31G level, leaving in total $16$ electrons in $36$ molecular orbitals of our effective description of the molecule. We relax the one-dimensional structure using the Hartree-Fock implementation of Molpro\cite{MOLPRO-WIREs}, resulting in the geometry $\overline{HC_1}=1.051389486 \text{\normalfont\AA}$, $\overline{C_1C_2}=1.196016909 \text{\normalfont\AA} $, $\overline{C_2I}=1.997857206 \text{\normalfont\AA}$.

The ionization in the experiment results in a strong hole density at the iodine atom. We shape our initial state accordingly and prepare the molecule in a super positional state of an ionization at the highest occupied molecular orbital (HOMO) and an ionization at the second highest molecular orbital (HOMO-1)
\begin{align}
   \ket{\psi(t_0)} = \frac{1}{\sqrt{2}} \left( \ket{HOMO} - \ket{HOMO-1} \right), \label{eq:iodoacetylene_initial}
\end{align}
where $\ket{HOMO}$ is the Hartree-Fock ground state with one electron removed from the HOMO orbital and $\ket{HOMO-1}$ is the Hartree-Fock ground state with one electron from the HOMO - 1 orbital. This state is not an energy eigenstate of the system and the electrons will evolve dynamics on the time scale of femtoseconds.

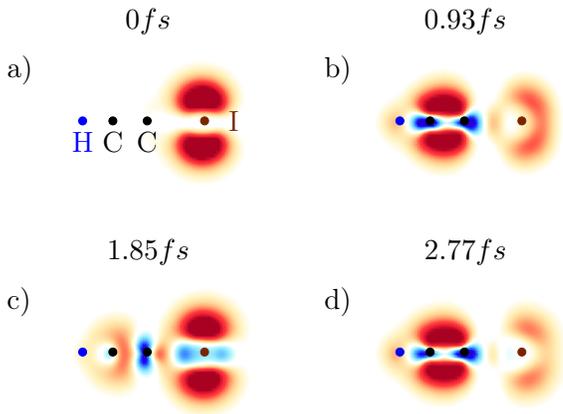
\begin{figure}
   \centering
   \begin{tikzpicture}
   \begin{groupplot}[
      group style={
          group name=my plots,
          group size=2 by 2,
          ylabels at=edge left,
          horizontal sep=5pt,
          vertical sep=30pt
      },
      footnotesize,
      width=160pt,
      tickpos=left,
      ytick align=outside,
      xtick align=outside,
      enlarge x limits=false 
  ]
  \nextgroupplot[
   title = {$0fs$},        
   ymax = 5,
   ymin = -5,
   xmin = -5,
   xmax = 15,
   axis equal image,
   point meta min=-0.05,
   point meta max=0.05,
   hide axis,
   colormap name=temp,
]
   \addplot [matrix plot*,point meta=explicit, shader=interp] file [meta=index 2] {plots/c2hi_density/density_t_000.dat};
   \draw[fill, color=blue] (axis cs:0.687832159,0) circle (1.5pt) node[below] {\small H};
   \draw[fill, color=black] (axis cs:2.674670195,0) circle (1.5pt) node[below] {\small C};
   \draw[fill, color=black] (axis cs:4.93481443,0) circle (1.5pt) node[below] {\small C};
   \draw[fill, color=Brown] (axis cs:8.710217113,0) circle (1.5pt) node[xshift=5pt,right] {\small I};
   \node[anchor=north west] at (rel axis cs:0,1) {\small a)};
   \nextgroupplot[
      title = {$0.93fs$},        
      ymax = 5,
      ymin = -5,
      xmin = -5,
      xmax = 15,
      axis equal image,
      point meta min=-0.05,
      point meta max=0.05,
      hide axis,
      colormap name=temp,
   ]
   \addplot [matrix plot*,point meta=explicit, shader=interp] file [meta=index 2] {plots/c2hi_density/density_t_093.dat};
   \draw[fill, color=blue] (axis cs:0.687832159,0) circle (1.5pt);
   \draw[fill, color=black] (axis cs:2.674670195,0) circle (1.5pt);
   \draw[fill, color=black] (axis cs:4.93481443,0) circle (1.5pt);
   \draw[fill, color=Brown] (axis cs:8.710217113,0) circle (1.5pt);
   \node[anchor=north west ] at (rel axis cs:0,1) {\small b)};
   
   \nextgroupplot[
      title = {$1.85fs$},        
      ymax = 5,
      ymin = -5,
      xmin = -5,
      xmax = 15,
      axis equal image,
      point meta min=-0.05,
      point meta max=0.05,
      hide axis,
      colormap name=temp,
   ]
   \addplot [matrix plot*,point meta=explicit, shader=interp] file [meta=index 2] {plots/c2hi_density/density_t_185.dat};
   \draw[fill, color=blue] (axis cs:0.687832159,0) circle (1.5pt);
   \draw[fill, color=black] (axis cs:2.674670195,0) circle (1.5pt);
   \draw[fill, color=black] (axis cs:4.93481443,0) circle (1.5pt);
   \draw[fill, color=Brown] (axis cs:8.710217113,0) circle (1.5pt);
   \node[anchor=north west] at (rel axis cs:0,1) {\small c)};
   
   \nextgroupplot[
            title = {$2.77fs$},        
            ymax = 5,
            ymin = -5,
            xmin = -5,
            xmax = 15,
            axis equal image,
            point meta min=-0.05,
            point meta max=0.05,
            hide axis,
            colormap name=temp,
         ]
   \addplot [matrix plot*,point meta=explicit, shader=interp] file [meta=index 2] {plots/c2hi_density/density_t_277.dat};
   \draw[fill, color=blue] (axis cs:0.687832159,0) circle (1.5pt);
   \draw[fill, color=black] (axis cs:2.674670195,0) circle (1.5pt);
   \draw[fill, color=black] (axis cs:4.93481443,0) circle (1.5pt);
   \draw[fill, color=Brown] (axis cs:8.710217113,0) circle (1.5pt);
   \node[anchor=north west] at (rel axis cs:0,1) {\small d)};
      
   \end{groupplot}
\end{tikzpicture}
   \caption{Hole density of the iodoacetylene molecule at four different points in time. We use an MPS bond dimension of $D=200$, a Krylov space dimension of $N_K=5$ and a time step size of $\Delta t = 1as$. a) shows the initial hole population at the iodine atom b) the hole has migrated to the acetylene after $0.93fs$ of free evolution c) after $1.85fs$ the hole migrated back to the iodine atom d) the hole is again located at the carbon atoms. }
   \label{fig:iodoacetylene_dynamics}
\end{figure}




In Figure \ref{fig:iodoacetylene_dynamics} we observe the same charge migration between the iodine atom and the carbon pair that was observed before\cite{Kraus790,Woerner:233644}. The initial hole density at the iodine migrates to the acetylene and back, almost exactly like it was reported in the experiment. We used a Krylov space dimension of $N_K = 5$ with orthogonalized basis vectors and a time step size of $\Delta t = 1as$. The maximum MPS bond dimension used in this calculation is $D=200$. These results are well converged in terms of Krylov space dimension, time step size, MPS bond dimension and the number of active orbitals included in the calculation. See the Supporting Information for an analysis of the necessary number of active orbitals and the MPS bond dimension. 

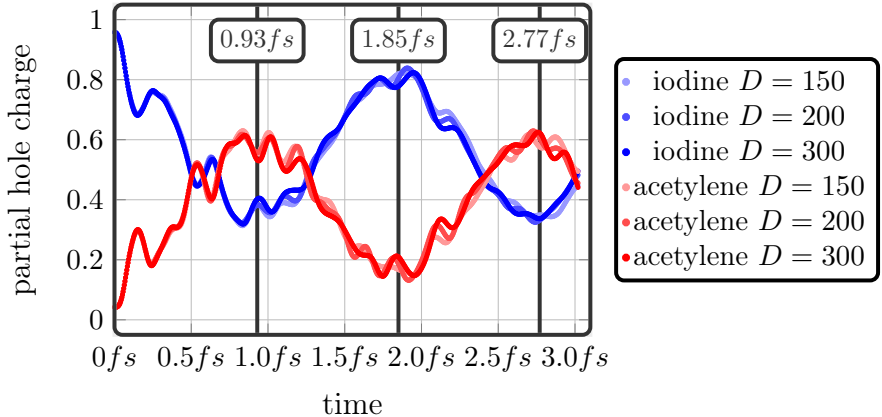
\begin{figure}
   \centering
   \begin{tikzpicture}
   \begin{axis}[
         xlabel={time},
         xlabel near ticks,
         width=225pt,
         height=170pt,
         xmin=0.0, 
         xmax= 3.1,
         xmajorgrids, 
         ymajorgrids, 
         xtick={0.0,0.5,1.0,1.5,2.0,2.5,3.0},
         xticklabels={$0fs$,$0.5fs$,$1.0fs$,$1.5fs$,$2.0fs$,$2.5fs$,$3.0fs$},
         ytick={0.0,0.2,0.4,0.6,0.8,1.0},
         ymin=-0.05,
         ymax=1.05,
         ylabel={partial hole charge},
         tick label style={font=\small},
         after end axis/.code={     \draw[line width = 1.5pt, color = gray!15] 
                                 (rel axis cs:0,0)rectangle(rel axis cs:1,1);
                                 \draw[line width = 1.5pt,rounded corners=3pt, color = black!80] 
                                 (rel axis cs:0,0)rectangle(rel axis cs:1,1);
                                    },
                                    legend entries={\small iodine $D=150$, \small iodine $D=200$, \small iodine $D=300$, \small acetylene $D = 150$, \small acetylene $D = 200$, \small acetylene $D = 300$},
                                    legend style={rounded corners=3pt, line width = 1.5pt,at={(1.05,0.5)},anchor=west,font=\small, /tikz/every even column/.append style={column sep=0.25cm}},  
         ]
         \addplot[only marks, scatter src=y, mark size=0.75pt, blue!40] table [x expr=\thisrowno{0} * 0.02418884254, y expr=6.67135515-\thisrowno{4} ] {plots/c2hi_part-charge/part-charge-low_N_26_150.dat};
         \addplot[only marks, scatter src=y, mark size=0.75pt, blue!70] table [x expr=\thisrowno{0} * 0.02418884254, y expr=6.67135515-\thisrowno{4} ] {plots/c2hi_part-charge/part-charge-low_N_26_200.dat};
         \addplot[only marks, scatter src=y, mark size=0.75pt, blue] table [x expr=\thisrowno{0} * 0.02418884254, y expr=6.67135515-\thisrowno{4} ] {plots/c2hi_part-charge/part-charge-low_N_26_300.dat};
         \addplot[only marks, scatter src=y, mark size=0.75pt, red!40 ] table [x expr=\thisrowno{0} * 0.02418884254, y expr=4.17651588+4.33514282-\thisrowno{2}-\thisrowno{3} ] {plots/c2hi_part-charge/part-charge-low_N_26_150.dat};   
         \addplot[only marks, scatter src=y, mark size=0.75pt, red!70 ] table [x expr=\thisrowno{0} * 0.02418884254, y expr=4.17651588+4.33514282-\thisrowno{2}-\thisrowno{3} ] {plots/c2hi_part-charge/part-charge-low_N_26_200.dat};
         \addplot[only marks, scatter src=y, mark size=0.75pt, red ] table [x expr=\thisrowno{0} * 0.02418884254, y expr=4.17651588+4.33514282-\thisrowno{2}-\thisrowno{3} ] {plots/c2hi_part-charge/part-charge-low_N_26_300.dat};
         \draw[line width = 1.5pt,color = black!80] (axis cs:0.93,-0.05) -- (axis cs:0.93,1.05) node[anchor=north,fill=white,rectangle,draw,rounded corners=3pt,yshift=-5pt, scale=0.9] {\small$0.93fs$};
         \draw[line width = 1.5pt,color = black!80] (axis cs:1.85,-0.05) -- (axis cs:1.85,1.05) node[anchor=north,fill=white,rectangle,draw,rounded corners=3pt,yshift=-5pt, scale=0.9] {\small$1.85fs$};
         \draw[line width = 1.5pt,color = black!80] (axis cs:2.77,-0.05) -- (axis cs:2.77,1.05) node[anchor=north,fill=white,rectangle,draw,rounded corners=3pt,yshift=-5pt, scale=0.9] {\small$2.77fs$};
   \end{axis}
\end{tikzpicture}
   \caption{The partial hole charge at the iodine atom and at the acetylene changing with time. The Krylov space dimension is $N_K=5$ and the time step size is $\Delta t = 1as$. Partial charges are displayed for various MPS bond dimensions, showing the good convergence. Reference points from experiment by Kraus et al.\cite{Kraus790} are highlighted, depicting transition points in the dynamics. The point $2.77fs$ was added for completeness. }
   \label{fig:iodoacetylene_part-charge}
\end{figure}

To quantify the charge migration dynamics, we calculate the hole population at the iodine atom and at the acetylene using L\"owdin population analysis\cite{doi:10.1063/1.1747632, doi:10.1002/qua.20981}. Thereby, we project the electron density onto the orthogonalized atomic orbitals of the atoms, allowing us to identify partial charges for each atomic species. To study migration involving the iodine, we distinguish between charge located at the iodine, as well as charge located at the two carbon atoms, namely the acetylene. It is clearly seen in Figure \ref{fig:iodoacetylene_part-charge} that the hole depopulates the iodine atom within the first femtosecond. Close to the time $t=0.93fs$ the depopulation reaches its maximum and the hole starts to migrate back to the iodine atom again. It takes another $0.92fs$ to populate the iodine atom, such that after $1.85fs$ we again have a maximum in the hole population at the iodine atom. This directly fits the experiment\cite{Kraus790}, while being in good agreement with time-dependent density functional theory calculations performed for this situation\cite{Woerner:233644}. Our results also agree with emission spectra\cite{F29777301406} for this system. 

All calculations with MPS bond dimension $D\geq150$ show this charge migration behavior. Our MPS approach is able to describe the many-electron state and to follow the correct dynamics for a few femtoseconds after ionization. This was obtained with MPS bond dimension as small as $D=150$, which is a significant reduction of the effective Hilbert space size. The corresponding full CI MPS bond dimension is $D_{FCI}= 458,681$ and therefore far from any calculation modern computers a capable of. This impressively shows the ability of the MPS approach to reduce the effective degrees of freedom in time-dependent quantum chemistry problems. Apart from the limited basis set and the core potentials, we obtained these results without any a-priori assumptions to the many-electron state. The effective Hilbert space is dynamically adjusted by the optimization of the time-dependent MPS.

\section{Conclusion}
This work benchmarked time evolution based on matrix product states in the context of charge migration. We have observed its potential in describing ultrafast dynamics on a full CI level. The fourth-order Runge-Kutta method, as well as the Krylov space time evolution method are both able to follow the full CI dynamics when calculating Green's functions. We compared results for a chain of ten hydrogen atoms and observed better behavior of the MPS based Krylov method compared to the RK4 implementation. Furthermore is the Krylov space method more flexible in the adaption to MPS. We showed that using larger Krylov space dimension does not necessarily increase the accuracy and that waiving the orthonormalization of the Krylov space basis vectors may increase the accuracy of the calculation when using MPS of limited bond dimension. We demonstrated this at the example of the water molecule that was singly ionized at the $1s$ orbital of the oxygen. The MPS of limited bond dimension need to be able to correspond to the desired state and constraints on the MPS may be counter productive. As new example, we studied charge migration in iodoacetylene and compared to recent literature. The MPS approach converged quickly with bond dimension allowing us to state that the presented dynamics are quasi full CI. The observed oscillation behavior corresponds directly to what has been reported earlier, with the observed oscillation frequency being within the time resolution of the experiment. 

The dynamic optimization of the included configurations and the rapid convergence with the MPS bond dimension allows for further studies of electron dynamics in molecules. We conclude this from the observed behavior of the shown examples. However, the general dependence of the MPS bond dimension and the time evolution parameters on the system properties still needs further systematic studies. In our calculations, we worked mostly in the static basis of molecular orbitals, which usually gives a good starting point, however, that might not be the optimal basis after a few femtoseconds. The time range of our MPS approach might be extended by adapting the orbital basis after some time, allowing an additional reduction of the MPS bond dimension. Such a dynamic optimization of the orbitals was successful for ground state studies{\cite{PhysRevLett.117.210402}} and will be implemented for time evolution in future work. 

The ideas and concepts disussed in this work are not limited to MPS. The same approach can be applied to more complex wave function decompositions such as tree tensor network states and even tensor networks in general. This will further improve handling of the emerging correlations in specific situations. Lastly, there is no need to keep the nuclei at fixed positions. Including moving nuclei will allow to study the transition from charge migration to charge transfer and even chemical reactions. The moving nuclei might even destroy electronic correlations\cite{PhysRevA.95.033425}, effectively decreasing the necessary MPS bond dimension capturing the many-electron state.

\begin{acknowledgement}
   D. Pfannkuche was supported by the excellence cluster \textit{The Hamburg Centre for Ultrafast Imaging Structure, Dynamics and Control of Matter at the Atomic Scale} (CUI, DFG-EXC1074) and L.-H. Frahm was supported by the \textit{PIER Helmholtz Graduate School}. The authors also like to thank R. Santra for valuable discussions on charge migration and C. Hubig for directions on the variational handling of matrix product states. Further we thank S. Wouters for providing additional documentation of the CheMPS2 code.
\end{acknowledgement}
   
\begin{suppinfo}
The Supporting Information is available free of charge on the ACS Publication website at DOI: 

    Convergence of time-dependent Green's function of the hydrogen chain; Convergence of the charge migration dynamics in iodoacetylene

\end{suppinfo}

   \tikzset{external/export next=false}

\begin{tocentry}
      \begin{tikzpicture}
         \node[inner sep=0pt] (t000) at (0,0) {\includegraphics[width=\linewidth]{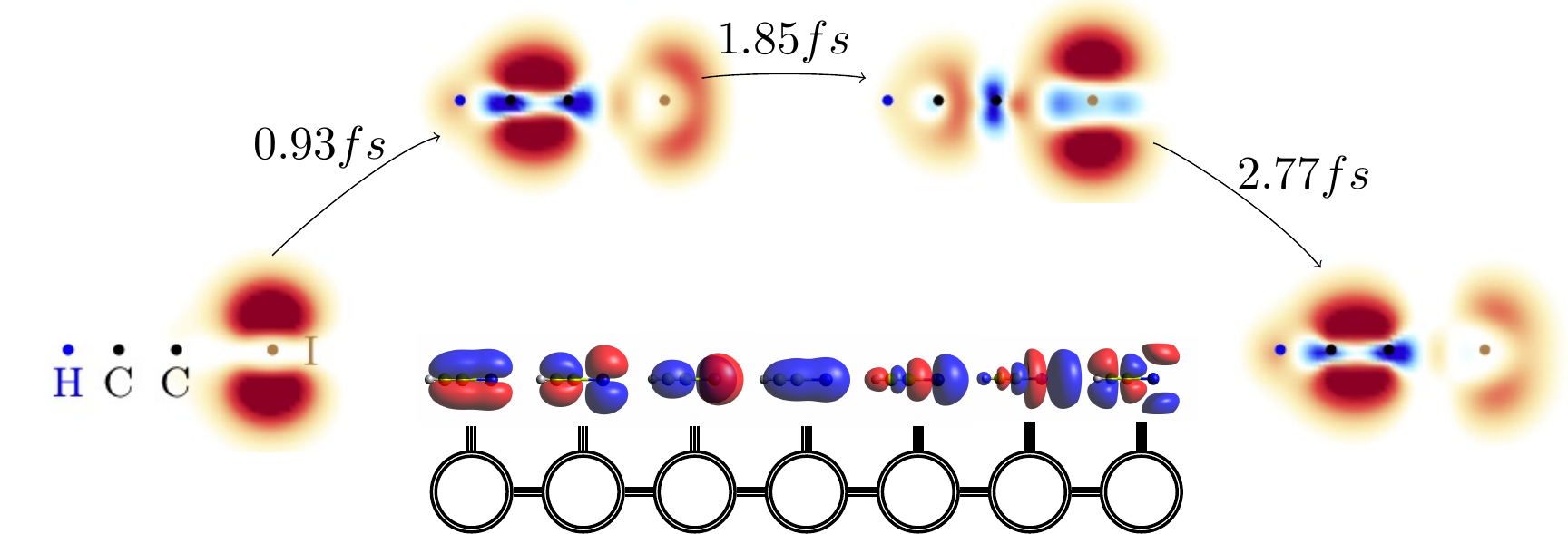}};
      \end{tikzpicture}
\end{tocentry} 

   \bibliography{manuscript} 
  
   \includepdf[pages=-]{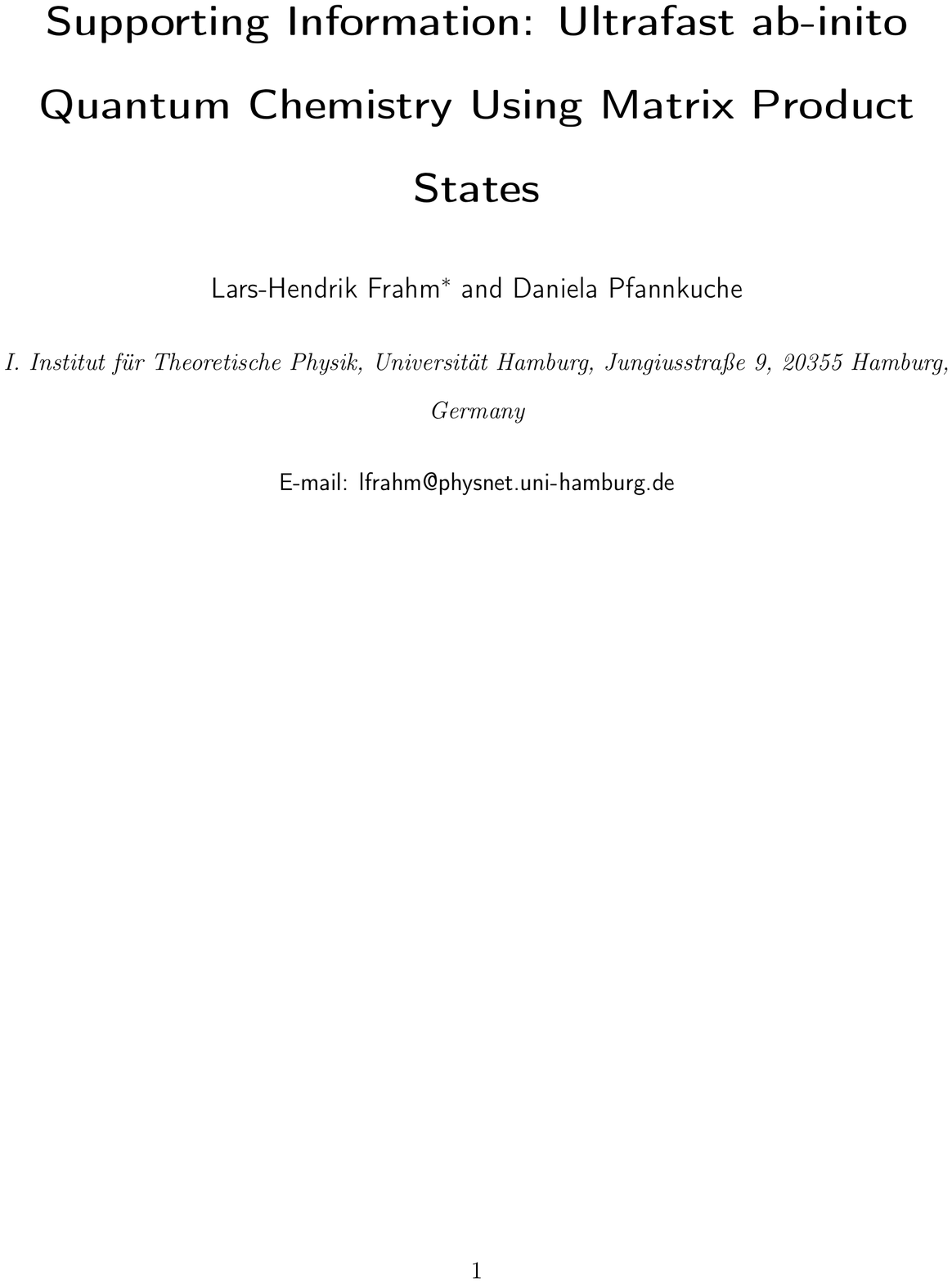}

\end{document}